\documentclass[preprint]{aastex}
\slugcomment{Accepted by the \emph{Astrophysical Journal}}

\shortauthors{Spergel et al.}

\shorttitle{WMAP First Year Results: Parameters}
\usepackage{epsfig}
\usepackage{graphicx}
\usepackage{amssymb}
\newcommand\tdf{2dFGRS}
\newcommand\mapext{WMAPext}
\newcommand\lya{Lyman $\alpha$\ }
\newcommand\lap{\la}

\newcommand\map{{\sl WMAP\ }}

\begin{document}
\title{First Year Wilkinson Microwave Anisotropy Probe ({\sl WMAP \/})
Observations: Determination of Cosmological Parameters}
\author{
D. N. Spergel \altaffilmark{2},
L. Verde \altaffilmark{2,3},
H. V. Peiris \altaffilmark{2},
E. Komatsu \altaffilmark{2},
M. R. Nolta \altaffilmark{4},
C. L. Bennett \altaffilmark{5},
M. Halpern  \altaffilmark{6},
G. Hinshaw \altaffilmark{5},
N. Jarosik \altaffilmark{4},
A. Kogut \altaffilmark{5},
M. Limon \altaffilmark{5,7},
S. S. Meyer \altaffilmark{8},
L. Page \altaffilmark{4},
G. S. Tucker \altaffilmark{5,7,9},
J. L. Weiland \altaffilmark{10},
E. Wollack \altaffilmark{5},
\& E. L. Wright \altaffilmark{11}}

\altaffiltext{1}{\map is the result of a partnership between Princeton
                 University and NASA's Goddard Space Flight Center. Scientific
                 guidance is provided by the \map\ Science Team.}
\altaffiltext{2}{Dept of Astrophysical Sciences, 
            Princeton University, Princeton, NJ 08544}
\altaffiltext{3}{Chandra Postdoctral Fellow}
\altaffiltext{4}{Dept. of Physics, Jadwin Hall, 
            Princeton, NJ 08544}
\altaffiltext{5}{Code 685, Goddard Space Flight Center, 
            Greenbelt, MD 20771}
\altaffiltext{6}{Dept. of Physics and Astronomy, University of 
            British Columbia, Vancouver, BC  Canada V6T 1Z1}
\altaffiltext{7}{National Research Council (NRC) Fellow}
\altaffiltext{8}{Depts. of Astrophysics and Physics, EFI and CfCP, 
            University of Chicago, Chicago, IL 60637}
\altaffiltext{9}{Dept. of Physics, Brown University, 
            Providence, RI 02912}
\altaffiltext{10}{Science Systems and Applications, Inc. (SSAI), 
            10210 Greenbelt Road, Suite 600 Lanham, Maryland 20706}
\altaffiltext{11}{UCLA Astronomy, PO Box 951562, Los Angeles, CA 90095-1562}

\email{dns@astro.princeton.edu}
\begin{abstract}
\map precision  data enables accurate testing
of cosmological models.  We find that 
the emerging standard
model of cosmology,
a flat $\Lambda-$dominated universe seeded by a nearly scale-invariant 
adiabatic Gaussian fluctuations, fits the \map data.   For the {\sl WMAP} data only,
the best fit parameters are
 \ensuremath{h = 0.72\pm 0.05},
 \ensuremath{\Omega_bh^2 = 0.024 \pm 0.001}, 
 \ensuremath{\Omega_mh^2 = 0.14 \pm 0.02}, 
 \ensuremath{\tau = 0.166^{+ 0.076}_{- 0.071}},  
 \ensuremath{n_s = 0.99 \pm 0.04}, and
 \ensuremath{\sigma_8 = 0.9 \pm 0.1}.  
 With
parameters fixed only by \map data, we can fit finer scale CMB measurements
and
measurements of large scale structure (galaxy surveys and the Lyman $\alpha$
forest).
This simple model is also consistent with a host of other astronomical 
measurements: its inferred age of the universe
is consistent with stellar ages, the baryon/photon ratio 
is consistent with measurements of the [D]/[H] ratio, and the inferred Hubble 
constant is consistent with local observations of the expansion rate.  
We then fit the model parameters to a combination of
\map data with other finer scale CMB
experiments (ACBAR and CBI), \tdf\ measurements
and \lya forest data to find the model's best fit cosmological
parameters:
 \ensuremath{h = 0.71^{+ 0.04}_{- 0.03}},
 \ensuremath{\Omega_bh^2 = 0.0224 \pm 0.0009}, 
 \ensuremath{\Omega_mh^2 = 0.135^{+ 0.008}_{- 0.009}}, 
 \ensuremath{\tau = 0.17 \pm 0.06},  $n_s$(0.05 Mpc$^{-1}) =$
 \ensuremath{0.93 \pm 0.03}, and
 \ensuremath{\sigma_8 = 0.84 \pm 0.04}.  
{\sl WMAP}'s best determination of $\tau=0.17 \pm 0.04$ 
arises directly from the TE data and not from this model fit, but they are 
consistent.
These parameters
imply that the age of the universe is
\ensuremath{13.7 \pm 0.2 \mbox{ Gyr}}.
With the Lyman $\alpha$ forest data, the model  favors but does not require a slowly varying
spectral index.  The significance of this running
index is sensitive to the uncertainties in the Lyman $\alpha$ forest.

By combining \map data with other astronomical data,
we constrain
the geometry of the universe: $\Omega_{tot} = 1.02 \pm 0.02$,
and 
the equation of state of the dark energy, $w < -0.78$ (95\%
confidence limit assuming $w \ge -1$.). 
The combination of \map and \tdf\ data constrains
the energy density in stable neutrinos:
$\Omega_\nu h^2 < 0.0076$ (95\% confidence limit).
For 3 degenerate neutrino species, this limit implies that their mass
is less than 0.23~eV (95\% confidence limit).  The \map detection of early reionization
rules out warm dark matter.
\end{abstract}

\keywords{
 cosmic microwave background --- cosmology: observations
 --- early universe
}

\section{INTRODUCTION}

Over the past century, a standard cosmological model has emerged: 
With relatively few parameters, the model 
describes the evolution of the Universe and 
astronomical observations on scales ranging from  a few to thousands of 
Megaparsecs. 
In this model the Universe is  spatially flat, homogeneous and 
isotropic on large scales, composed of radiation,
ordinary matter (electrons, protons, neutrons and neutrinos), 
non-baryonic cold dark matter, and dark energy.  
Galaxies and large-scale structure grew gravitationally from 
tiny, nearly scale-invariant adiabatic Gaussian fluctuations.  
The Wilkinson Microwave Anisotropy Probe (\map)
data offer a demanding quantitative test of this model.

The {\sl WMAP} data are powerful because they result from a mission that 
was carefully designed to limit systematic measurement errors 
\citep{bennett/etal:2003, bennett/etal:2003b, hinshaw/etal:2003}.  A 
critical element of this design includes differential measurements of the 
full sky with a complex sky scan pattern.  The nearly uncorrelated noise 
between pairs of pixels, the accurate in-flight determination of the beam 
patterns \citep{page/etal:2003, page/etal:2003b, barnes/etal:2003}, and 
the  well-understood properties of the radiometers 
\citep{jarosik/etal:2003, jarosik/etal:2003b} are invaluable for this
analysis.

Our basic approach in this analysis is to begin by identifying the simplest
model that fits the \map data and determine the best fit parameters
for this model using \map data only without the use of
any significant priors on parameter values.  We then compare
the predictions of this model to other data sets and
find that the model is basically consistent with these data sets.
We then fit to combinations of the \map data and other
astronomical data sets and find the best fit global model.
Finally, we place constraints on alternatives to this model.

We begin by outlining our methodology (\S 2).  \citet{verde/etal:2003}
describes the details of the approach used here to compare theoretical
predictions of cosmological models to data.  In \S 3,
we fit a simple, six parameter $\Lambda$CDM model to the \map data-set (temperature-temperature  and temperature-polarization angular power spectra). 
In \S 4 we show that this simple model provides an acceptable fit not only to the \map data, but also to a host of astronomical data.  We use the comparison with these other datasets to test the validity of the model rather than further constrain the model parameters.
In \S 5,
we include large scale structure data from the 2dF
Galaxy Redshift Survey (2dFGRS, \citet{colless/etal:2001}) and Lyman $\alpha$ forest data to perform a joint likelihood analysis for the cosmological parameters.  We find that the data favors a slowly varying
spectral index. This seven parameter model is our best fit to the full data 
set.  In \S 6, we relax some of the  minimal assumptions of the model 
by adding extra parameters to the model.
We examine non-flat models, dark energy models in which the properties 
of the dark energy are 
parameterized by an effective equation of state, and models
with gravity waves. By adding extra parameters  we introduce  degenerate sets of models consistent 
with the \map data alone. We lift these degeneracies  by including 
additional microwave background data-sets (CBI, ACBAR) and observations of large-scale structure.  We use these combined data sets to 
place strong limits on the geometry of the universe, the neutrino mass,
the energy density in gravity waves, and the properties of the dark energy. In \S 7, we note an intriguing discrepancy between the standard model 
and the \map data on the largest angular scales and speculate on its 
origin.  In \S 8, we conclude and present parameters for our best fit model.

\section{BAYESIAN ANALYSIS OF COSMOLOGICAL DATA}

The basic approach of this paper is to find the simplest model consistent with cosmological data. We begin by fitting a simple six parameter model first to the \map data and then to other cosmological data sets.  We then consider more complex cosmological models and evaluate whether they are a better description of the cosmological data. 
Since \citet{komatsu/etal:2003} found no evidence
for non-Gaussianity in the \map data, we assume  the primordial
fluctuations are Gaussian random phase throughout this paper.
For each model studied in the paper, we use a Monte Carlo Markov Chain 
to explore the likelihood surface. We assume flat priors in our basic 
parameters, impose positivity
constraints on the matter and baryon density (these limits lie at such low
likelihood that they are unimportant for the models.  We assume a flat prior in $\tau$, the optical
depth, but bound $\tau < 0.3$.  This
prior has little effect on the fits but keeps the Markov Chain
out of unphysical regions of parameter space.
For each model, we determine the best fit parameters  from the peak of the 
N-dimensional likelihood surface. For each parameter in the model we also compute its one dimensional likelihood function by marginalizing over all other parameters; we then quote the (1-dimensional)  expectation value\footnote{In a   Monte Carlo Markov Chain, it is a more robust quantity  than the mode of the a posteriori marginalized distribution.} as our best estimate for the parameter:
\begin{equation}
\langle \alpha_i\rangle = \int d^N\alpha {\cal L}({\bf \alpha}) \alpha_i,
\end{equation}
where $\vec{\alpha}$ denotes a point in the N-dimensional parameter space  (in our application these are points --sets of cosmological parameters--  in the output of the Markov Chain), ${\cal L}$ denotes the likelihood (in our application the ``weight'' given by the chain to each point).
The \map temperature (TT) angular power spectrum and the \map temperature-polarization (TE) angular power spectrum are our core data sets for
the likelihood analysis.  \citet{hinshaw/etal:2003} and \citet{kogut/etal:2003} describe how to obtain the temperature and temperature-polarization angular
power spectra respectively from the maps. \citet{verde/etal:2003} describes our basic methodology for evaluating
the likelihood functions using a Monte Carlo Markov Chain algorithm and
for including data-sets other than \map in our analysis.
In addition to \map data we use recent results from the CBI \citep{pearson/etal:2002}
and ACBAR \citep{kuo/etal:2002} experiments. We also
use the \tdf \ measurements of the power spectrum \citep{percival/etal:2001} and
the bias parameter \citep{verde/etal:2002},
measurements of the Lyman $\alpha$ power spectrum \citep{croft/etal:2002,
gnedin/hamilton:2002}, supernova Ia 
measurements of the angular diameter distance relation
\citep{garnavich/etal:1998,riess/etal:2001}, and the Hubble Space Telescope Key Project
measurements of the local expansion rate of the universe
\citep{freedman/etal:2001}.

\section{POWER LAW $\Lambda$CDM MODEL AND THE \map DATA}
We begin by considering a basic cosmological model: a flat Universe  with  radiation, baryons, cold dark matter and cosmological constant, and a power-law  power spectrum of adiabatic primordial fluctuations. 
As we will see, this model does a remarkably good job of describing \map TT and TE power spectra  with only six parameters:
the Hubble constant $h$ (in units of 100 km/s/Mpc),  the physical matter and baryon  densities $w_m\equiv\Omega_mh^2$ and $w_b\equiv\Omega_bh^2$, the optical depth to the  decoupling surface, $\tau$, the scalar spectral index $n_s$ and
$A$, the normalization
parameter in the CMBFAST code version 4.1 with option UNNORM.
\citet{verde/etal:2003} discusses the relationship between $A$ and
the amplitude of  curvature
fluctuations at horizon crossing, $|\Delta R|^2 =
2.95 \times 10^{-9} A$.
In \S 4, we show that
this model is also in acceptable  agreement 
with a wide range of astronomical data.

\begin{figure}
\figurenum{1}
\epsscale{0.6}
\rotatebox{90}{\plotone{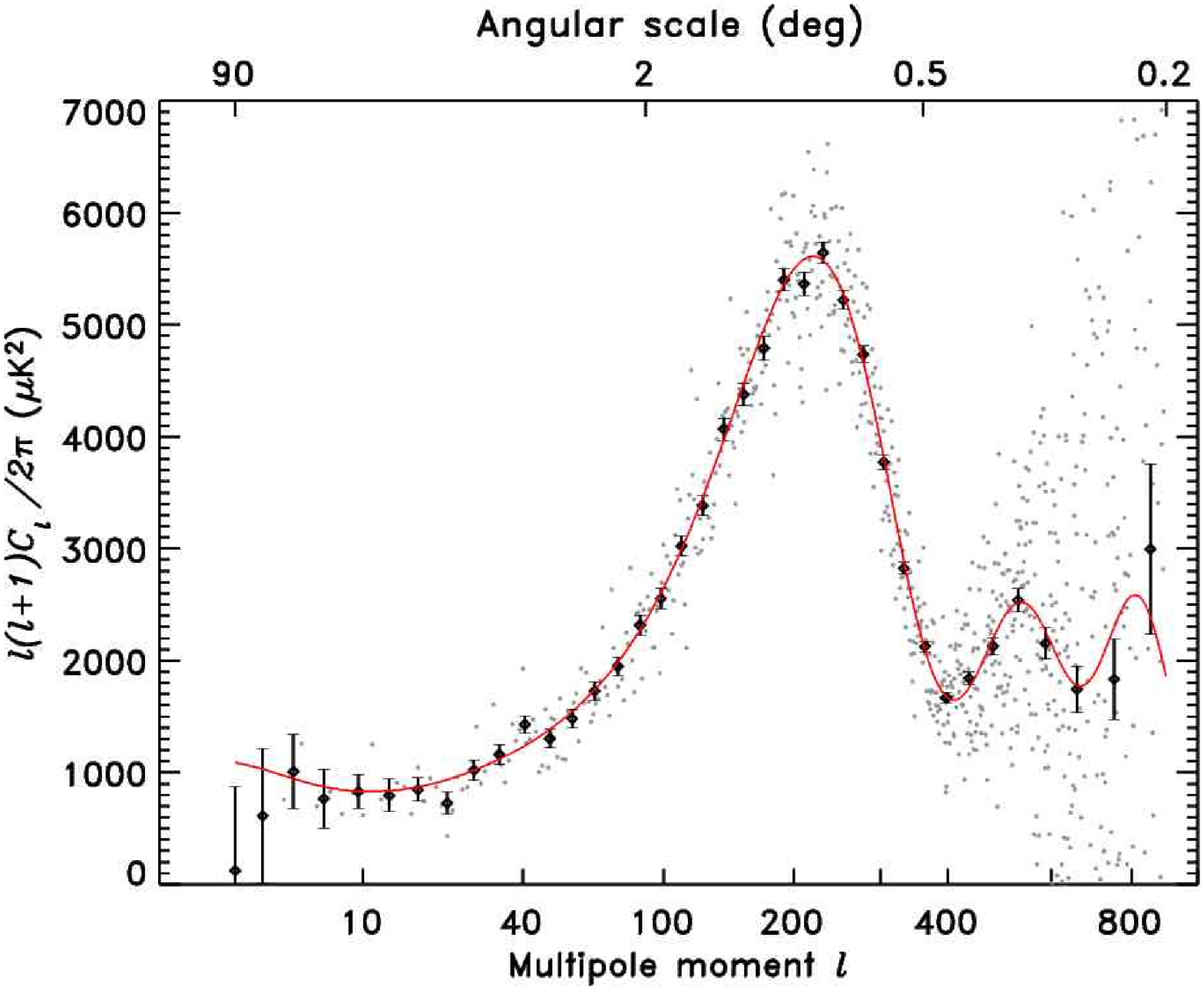}}
\caption{This figure compares the best fit power
law $\Lambda$CDM  model to the \map temperature angular power 
spectrum.  The gray dots are the unbinned data.
\label{fig:clfit}}
\end{figure}

\begin{figure}
\figurenum{2}
\epsscale{0.6}
\rotatebox{90}{\plotone{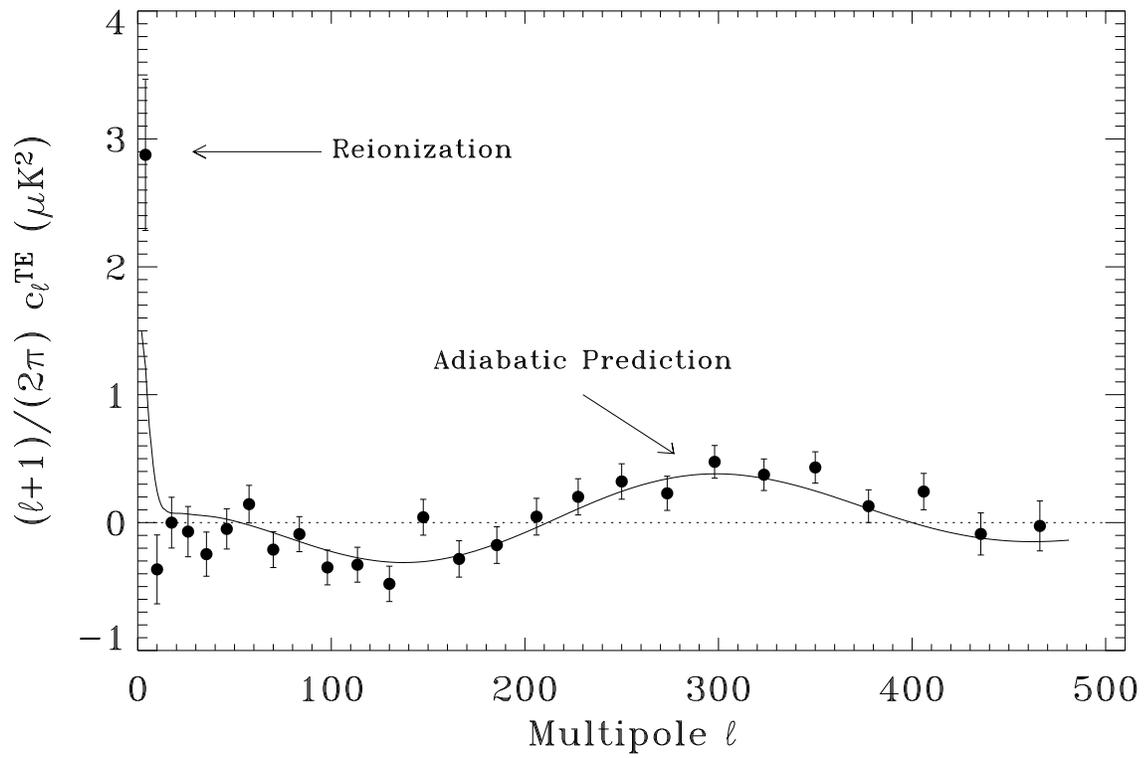}}
\caption{This figure compares the best fit power
law $\Lambda$CDM  model to the \map TE angular power 
spectrum. 
\label{fig:te_fit}}
\end{figure}

\begin{figure}
\figurenum{4}
\plotone{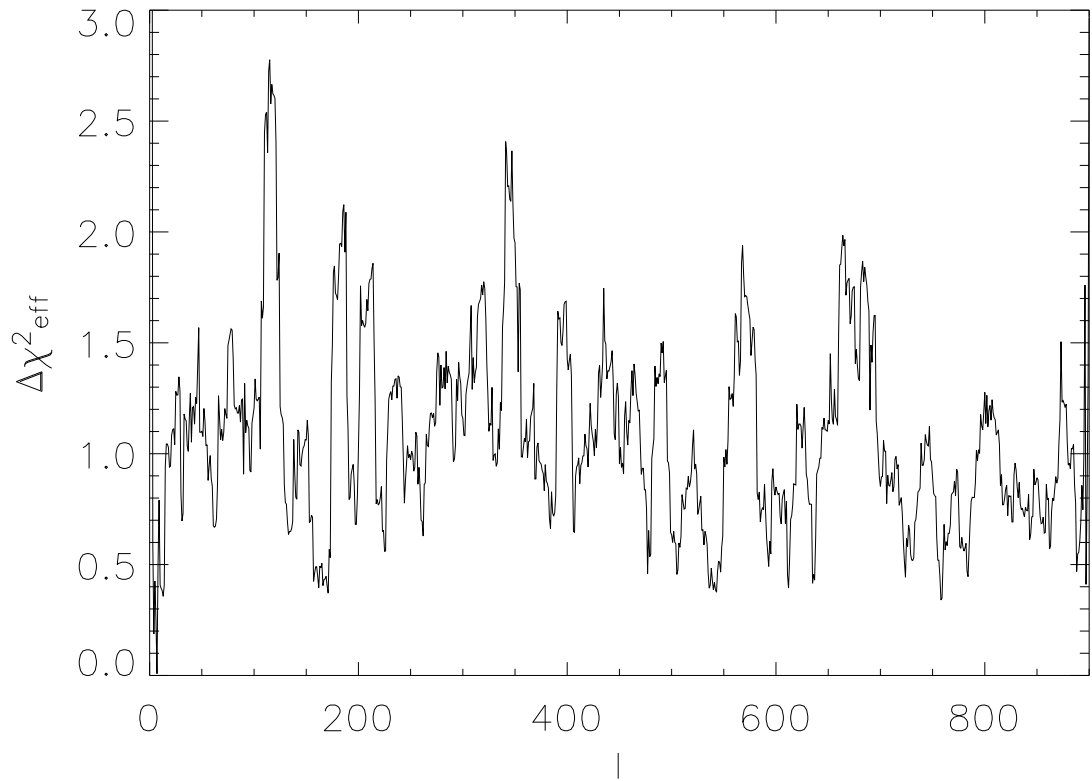}
\caption{This plot shows the  contribution to $2 \ln {\cal L}$ per multipole binned at $\Delta l = 15$. 
The excess $\chi^2$ comes primarily from three regions, one
around $\ell \sim 120$, one around $\ell\sim  200$ and the other around $\ell \sim 340$.
\label{fig:res_fit}}
\end{figure}

This simple model provides an acceptable fit to both the \map TT and TE data 
(see Figure \ref{fig:clfit} and \ref{fig:te_fit}).  
The reduced\footnote{Here, $\chi^2_{eff}\equiv-2\ln {\cal L}$ and  $\nu$
is number of data minus the number of parameters.  
We have used 100,000 Monte Carlo realization of the \map data with our 
mask, noise and angle-averaged beams and found that the 
$\left\langle-2\ln {\cal L}/\nu\right\rangle = 1$ 
for the simulated temperature data.} 
$\chi_{eff}^2$ for the full fit is 1.066 for 1342 degrees of freedom,
which has a probability of  $\sim 5$\%.
For the TT data alone, $\chi_{eff}^2/\nu = 1.09$, which for 
893 degrees of freedom has a probability of 3\%. 
Most of the excess $\chi_{eff}^2$ is due to the inability
of the model to fit sharp features in the power spectrum near $l \sim 120$,
the first TT peak
and at $l\sim 350$. In Figure \ref{fig:res_fit} we show the  contribution to 
$\chi_{eff}^2$ per multipole.  The overall excess variance is likely due to our not including several
effects, each  contributing roughly $0.5-1\%$ to our power spectrum covariance
near the first peak and trough: gravitational lensing of the CMB 
\citep{hu:2001}, the spatial variations in the effective
beam of the \map experiment 
due to variations in our scan orientation between
 the ecliptic pole and plane regions
\citep{page/etal:2003b,hinshaw/etal:2003b}, and non-Gaussianity in the noise
maps due to the $1/f$ striping.
Including these effects would increase our estimate of
the power spectrum uncertainties and improve our estimate 
of $\chi_{eff}^2$.  Our
next data release will include the corrections and errors associated with
the beam asymmetries. 
The features in the measured 
power spectrum could be due to underlying
features in the primordial power spectrum (see \S 5 of \citet{peiris/etal:2003}), but we do not yet attach cosmological significance to them.

\begin{deluxetable} {llll}
\tablecaption{Power Law $\Lambda$CDM Model Parameters-
\map Data Only
\label{tab:basic}}
\tablewidth{0pt}
\tablehead{
\colhead{Parameter} &   & Mean (68\% confidence range) & Maximum Likelihood }
\startdata
Baryon Density &$\Omega_b h^2$ &\ensuremath{0.024 \pm 0.001} &
 0.023\\
Matter Density & $\Omega_m h^2$ &\ensuremath{0.14 \pm 0.02} &
0.13 \\
Hubble Constant &$h$ &\ensuremath{0.72 \pm 0.05} 
&0.68 \\
Amplitude &$A$ &\ensuremath{0.9 \pm 0.1} & 0.78 \\
Optical Depth &$\tau$ &\ensuremath{0.166^{+ 0.076}_{- 0.071}} & 0.10 \\
Spectral Index &$n_s$ &\ensuremath{0.99 \pm 0.04}  & 0.97 \\
&$\chi^2_{eff}/\nu$ &   &1431/1342\\
\enddata
\tablenotetext{a}{Fit to \map data only}
\end{deluxetable}

\begin{figure}
\figurenum{3}
\epsscale{1.0}
\plotone{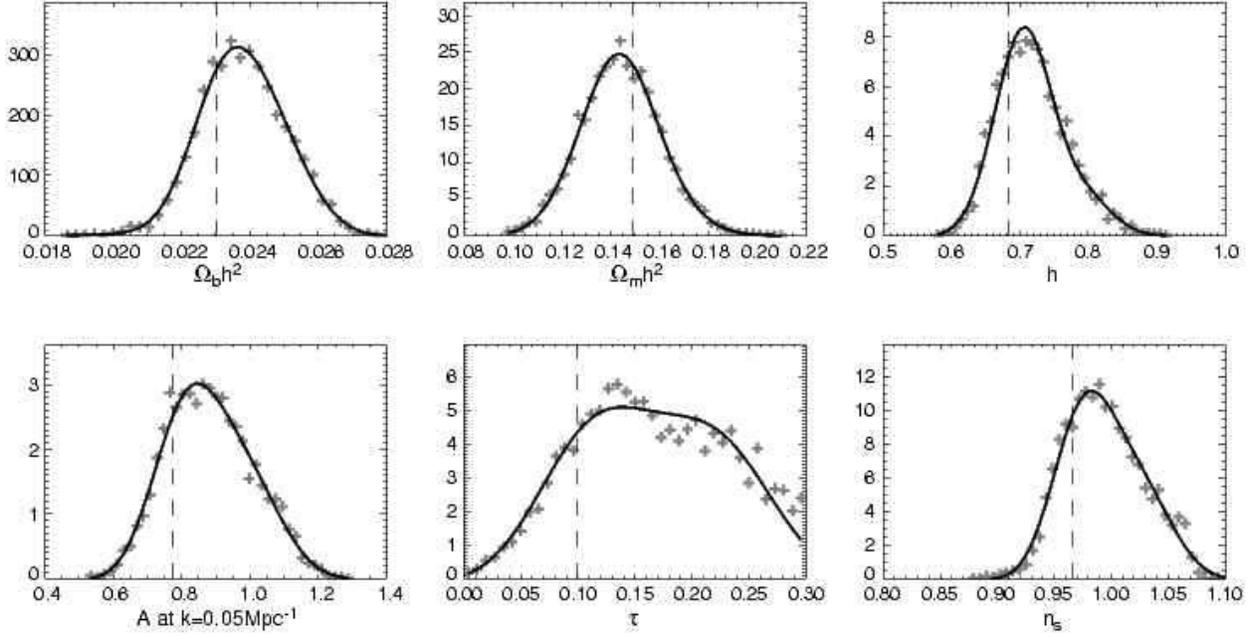}
\caption{This figure shows the likelihood function of the 
\map TT + TE data as a function of the basic parameters in the 
power law $\Lambda$CDM \map model. 
($\Omega_b h^2$, $\Omega_m h^2$, $h$, $A$, $n_s$ and $\tau$.) 
The points are the binned marginalized likelihood from the
Markov chain and the solid curve is an Edgeworth expansion of the Markov
 chains points.
The marginalized likelihood function is
nearly Gaussian for all of the parameters except for $\tau$.
The dashed lines show the maximum likelihood values of the global
six dimensional fit.  Since the peak in the likelihood, $x_{ML}$
is not the same as the expectation value of the likelihood
function, $<x>$, the dashed line does not lie at
the center of the projected likelihood.
\label{fig:lcdmmap}}
\end{figure}

\begin{figure}
\figurenum{5}
\plottwo{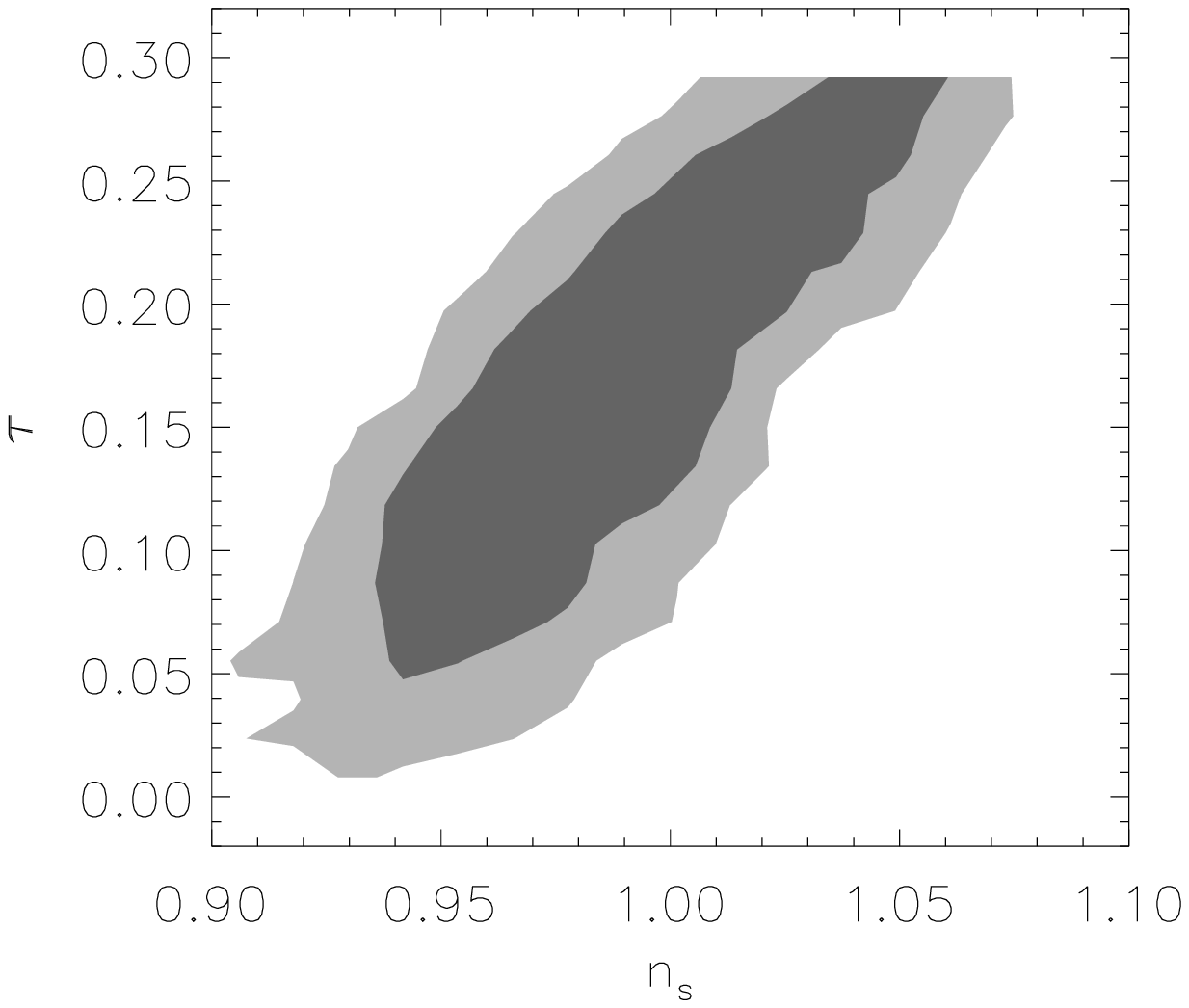}{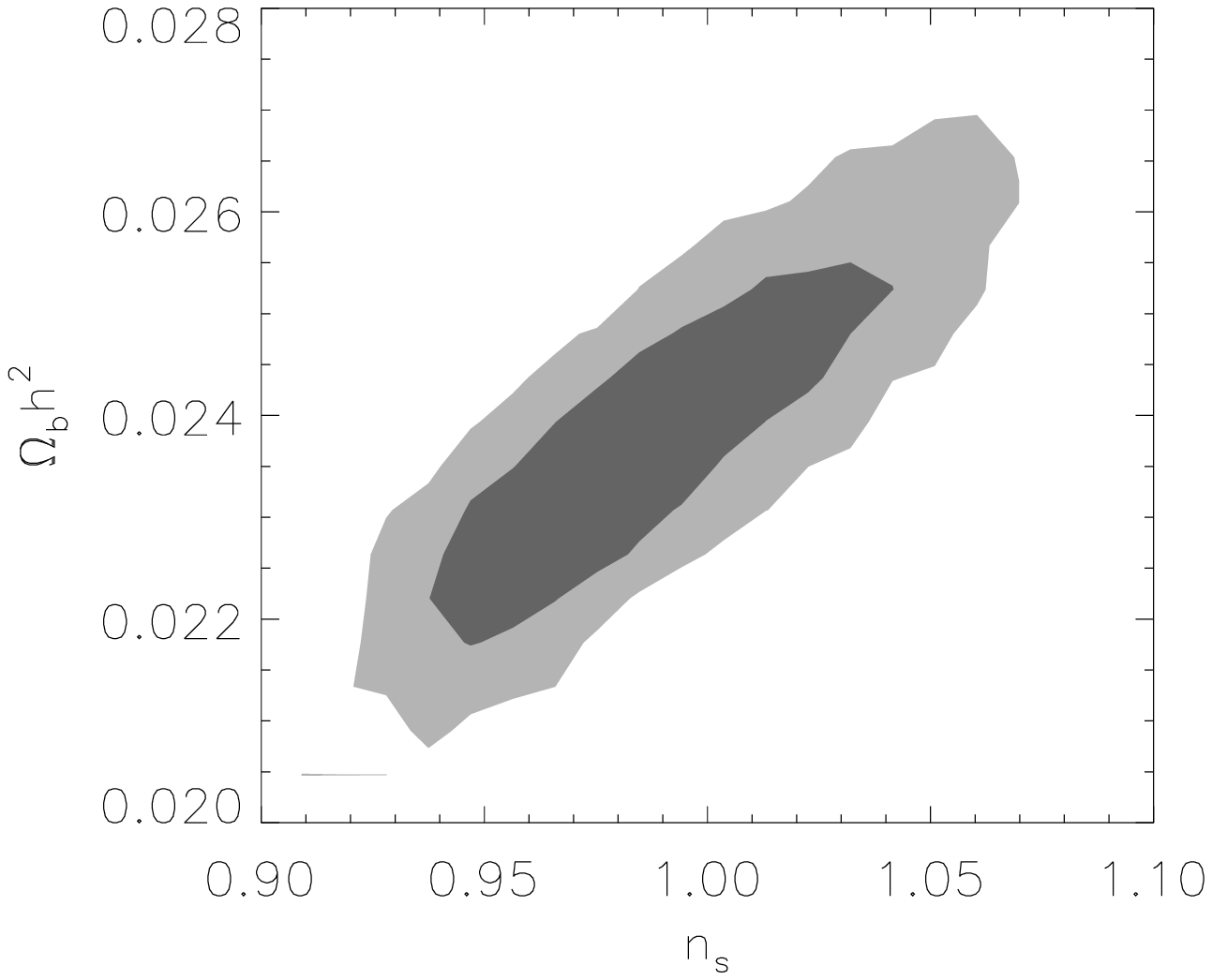}
\caption{Spectral Index Constraints. 
Left panel: the $n_s-\tau$ degeneracy in the \map data for a power-law 
$\Lambda$CDM model.  The TE observations constrain the value of $\tau$ 
and the shape of the $C_l^{TT}$ spectrum constrain a combination of $n_s$
and $\tau$. Right panel: $n_s-\Omega_b h^2$ degeneracy. 
The shaded regions show the joint one and two sigma confidence regions.
\label{fig:ns}}
\end{figure}

Table 1 lists the best fit parameters using  the \map data
alone for this model and Figure (\ref{fig:lcdmmap}) shows the marginalized
probabilities for each of the basic parameters in the model. The
values in the second column of
Table \ref{tab:basic} (and the subsequent parameter tables) are
expectation values for the marginalized distribution of each parameter
and the errors are the 68\% confidence interval.   The values in
the third column are the values at the peak of the likelihood
function.  Since we are projecting a high dimensional likelihood
function, the peak of the likelihood is not the same as the
expectation value of a parameter.
Most of the basic parameters are remarkably well determined within 
the context of this model.  
Our most significant parameter degeneracy (see Figure \ref{fig:ns})
is a degeneracy between $n_s$ and $\tau$.
The TE data favors $\tau \sim 0.17$ \citep{kogut/etal:2003};
on the other hand, the low value of the quadrupole  (see Figure
 \ref{fig:clfit} and \S \ref{sec:discrepancy}) and
the relatively low amplitude of fluctuations for $l < 10$
disfavors high $\tau$ as
reionization produces additional large scale anisotropies.  Because
of the combination of these two effects, the likelihood
surface is quite flat at its peak: the likelihood changes by
only $0.05$ as $\tau$ changes from $0.11 - 0.19$.  This particular
shape
depends upon the assumed form of the power spectrum: in \S \ref{sec:running},
we show that models with a scale-dependent spectral index have
a narrower likelihood function that is more centered around $\tau = 0.17$.

Since the \map data allows us to accurately determine many of the basic cosmological parameters, we can now infer a number of important derived quantities to very high accuracy; we do this by computing these quantities for each model in the MCMC and use the chain to determine their expectation values and  uncertainties.

\begin{deluxetable}{ll}
\tablecaption{Derived Cosmological Parameters
\label{table:derived}}
\tablewidth{0pt}
\tablehead{
\colhead{Parameter} &  \colhead{Mean (68\% confidence range)}}
\startdata
Amplitude of Galaxy Fluctuations &
\ensuremath{\sigma_8 = 0.9 \pm 0.1}\\
Characteristic Amplitude of Velocity Fluctuations &
\ensuremath{\sigma_8\Omega_m^{0.6} = 0.44 \pm 0.10}\\
Baryon Density/Critical Density &
\ensuremath{\Omega_b = 0.047 \pm 0.006}\\
Matter Density/Critical Density &
\ensuremath{\Omega_m = 0.29 \pm 0.07}\\
Age of the Universe &
\ensuremath{t_0 = 13.4 \pm 0.3 \mbox{ Gyr}}\\
Redshift of Reionization\tablenotemark{b}&
\ensuremath{z_r = 17 \pm 5}\\
Redshift at Decoupling &
\ensuremath{z_{dec} = 1088^{+ 1}_{- 2}}\\
Age of the Universe at Decoupling &
\ensuremath{t_{dec} = 372 \pm 14 \mbox{ kyr}} \\
Thickness of Surface of Last Scatter &
\ensuremath{\Delta z_{dec} = 194 \pm 2}\\
Thickness of Surface of Last Scatter &
\ensuremath{\Delta t_{dec} = 115 \pm 5 \mbox{ kyr}} \\
Redshift at Matter/Radiation Equality &
\ensuremath{z_{eq} = 3454^{+ 385}_{- 392}}\\
Sound Horizon at Decoupling &
\ensuremath{r_s = 144 \pm 4 \mbox{ Mpc}}\\
Angular Diameter Distance to the Decoupling Surface &
\ensuremath{d_A = 13.7 \pm 0.5 \mbox{ Gpc}}\\
Acoustic Angular Scale\tablenotemark{c} &
\ensuremath{\ell_A = 299 \pm 2}\\
Current Density of Baryons &
\ensuremath{n_b = (2.7 \pm 0.1) \times 10^{-7} \mbox{ cm$^{-3}$}}\\
Baryon/Photon Ratio &
\ensuremath{\eta = (6.5^{+ 0.4}_{- 0.3}) \times 10^{-10} \mbox{ }}\\
\enddata
\tablenotetext{a}{Fit  to the \map data only}
\tablenotetext{b}{Assumes ionization fraction, $x_e = 1$}
\tablenotetext{c}{ $l_A=\pi d_C/r_s$}
\end{deluxetable}

\clearpage

Table \ref{table:derived} lists  cosmological parameters based on
fitting a power law (PL) CDM model to the \map data only.
The parameters $t_{dec}$ and $z_{dec}$ are determined by using
 the CMBFAST code \citep{seljak/zaldarriaga:1996} to compute the redshift of
 the CMB ``photosphere'' 
(the peak in the photon visibility function). We determine the thickness
of the decoupling surface by measuring $\Delta z_{dec}$ and $\Delta
t_{dec}$, the full-width at half maximum of the visibility function. The
age of the Universe is derived by integrating the Friedmann equation, and
$\sigma_8$ 
(the linear theory
predictions for the amplitude of fluctuations within 8 Mpc/h spheres)
from the linear matter power spectrum at $z=0$ is computed by CMBFAST. 

\section{COMPARSION WITH ASTRONOMICAL PREDICTIONS}
In this section, we compare the predictions of the best fit
power law $\Lambda$CDM model to other cosmological observations.   We
also list in Table \ref{tab:derived_best} the best fit model to the full data set: a $\Lambda$CDM
model with a running spectral index (see \S \ref{sec:running}).
In particular we consider determinations of the local expansion rate (i.e. the Hubble constant), the amplitude of fluctuations on galaxy scales, the baryon abundance, ages of the oldest stars, large scale structure data and supernova Ia data. We also consider if our determination of the  reionization redshift is consistent with the prediction for structure formation in our best fit Universe and with recent
models of reionization.
In \S 5 and 6, we add some
of these data sets to the \map data to better constrain parameters and cosmological models.

\subsection{Hubble Constant}

CMB observations do not directly measure the local expansion rate of the Universe rather they measure the conformal distance to the decoupling surface 
and the matter-radiation ratio through the amplitude of the early 
Integrated  Sachs Wolfe (ISW) contribution relative to the height of the first peak. For our power law
$\Lambda$CDM model, this is enough information to ``predict'' the local expansion rate.  Thus, local Hubble constant measurements are an important test of
our basic model. 

The Hubble Key Project \citep{freedman/etal:2001} has carried out an extensive program of using Cepheids to calibrate several different secondary distance indicators (Type Ia supernovae,
Tully-Fisher, Type II supernovae, and surface brightness fluctuations).
With a distance modulus of $18.5$ for the LMC,
their combined estimate for the Hubble constant
is $ H_0=72 \pm 3 ({\rm stat.}) \pm 7 ({\rm systematic})$~km/s/Mpc.  
The agreement between the HST Key Project
value and our value, \ensuremath{h = 0.72 \pm 0.05}, is striking,
given that the two methods rely on different observables, different underlying
physics, and different model assumptions.   

As we will show in \S 6, models with equation of state for the 
dark energy very different from  a cosmological constant (i.e., $w=-1$) 
only fit the \map data if the Hubble constant is much smaller than 
the Hubble Key Project  value.
An independent determination of the Hubble constant that makes different
assumptions than the traditional distance ladder can be obtained by
combining Sunyaev-Zel'dovich and X-ray flux measurements of clusters of
galaxies, under the assumption of sphericity for the density and temperature profile
of clusters.
This method is sensitive to the Hubble constant at intermediate redshifts ($z \sim 0.5$), rather than in the nearby universe.
\citet{reese/etal:2002}, \citet{jones/etal:2001}, and \citet{mason/etal:2001} have obtained values for the Hubble constant systematically smaller than, the Hubble Key Project  and \map $\Lambda$CDM  model  determinations, but
all consistent at the  $1 \sigma$ level.
Table (\ref{tab:h}) summarizes recent Hubble constant determinations and compares them with the \map  $\Lambda$CDM  model value.

\subsection{Amplitude of Fluctuations}

The overall amplitude of fluctuations on large-scale structure scales has been recently determined from weak lensing surveys, clusters number counts and 
peculiar velocities from galaxy surveys.
Weak lensing surveys and peculiar velocity measurements are most sensitive to the combination $\sigma_8\Omega_m^{0.6}$, cluster abundance at low redshift is sensitive to a very similar parameter combination $\sigma_8\Omega_m^{0.5}$, but counts
of high redshift clusters can break the degeneracy. 

\subsubsection{Weak Lensing} 

Weak lensing directly probes the amplitude of mass fluctuations along the
line of sight to the background galaxies. Once the redshift distribution
of the background galaxies is  known, this technique directly probes
gravitational potential fluctuations, and therefore can be easily compared with our CMB model predictions
for the amplitude of dark matter fluctuations. Several groups have reported weak shear measurements within the past year (see Table \ref{tab:a} and \citet{vanwaerbeke/etal:2002a} for recent review):  while there is significant scatter in the reported
amplitude, the best fit model to the \map data lies
in the middle of the reported range. As these shear measurements
continue to improve, the combination of \map observations and lensing measurements will be a powerful probe of cosmological models.

\begin{deluxetable} {lll}
\tablecaption{Recent Hubble Constant Determinations
\label{tab:h}}
\tablehead{
\colhead{Method} &  \colhead{Mean (68\% confidence range)}
& \colhead{Reference}}
\startdata
Hubble Key Project & $72 \pm 3 \pm 7$& \citet{freedman/etal:2001} \\
SZE + X-ray & $60 \pm 4_{-18}^{+13} $&\citet{reese/etal:2002}\\
            & $66_{-11}^{+14}\pm 15$ & \citet{mason/etal:2001}\\ 
\map PL $\Lambda$CDM model & 
$72 \pm 5$
& \S 3 \\
\enddata
\end{deluxetable}

\begin{deluxetable}{lll}
\tablecaption{Amplitude of Fluctuations, $\sigma_8$ \label{tab:a}}
\tablehead{
\colhead{Method} &  \colhead{Mean (68\% confidence range)}
& \colhead{Reference}}
\startdata
PL $\Lambda$CDM + \map & $\ensuremath{0.9 \pm 0.1}$& \S 3 \\
Weak Lensing\tablenotemark{a,b} &$0.72 \pm 0.18$&\citet{brown/etal:2002}\\
            & $0.86^{+0.10}_{-0.09}$ &\citet{hoekstra/etal:2002}\\
            & $0.69^{+0.12}_{-0.16}$ & \citet{jarvis/etal:2002}\\ 
            & $0.96 \pm 0.12$ & \citet{bacon/etal:2002}\\
            & $0.92 \pm 0.2$ & \citet{refregier/rhodes/groth:2002} \\
            & $0.98 \pm 0.12$ & \citet{vanwaerbeke/etal:2002b} \\
Galaxy Velocity Fields\tablenotemark{b}
 & $0.73 \pm 0.1$ & \citet{willick/strauss:1998} \\
CBI SZ detection & $1.04 \pm 0.12$\tablenotemark{c}
 & \citet{komatsu/seljak:2002} \\
High redshift clusters\tablenotemark{b}
 & $0.95 \pm 0.1$ &\citet{bahcall/bode:2002} \\
\enddata
\tablenotetext{a}
{Since most weak lensing papers report 95\% confidence limits
in their papers, the table
lists the 95\% confidence limit for these experiments.}  \
\tablenotetext{b}{All of the $\sigma_8$ measurements
have been normalized to $\Omega_{m} = 0.287$, the best fit
value for a fit to the \map data only.}
\tablenotetext{c}{95\% confidence limit}
\end{deluxetable}

\subsubsection{Galaxy velocity fields}

The galaxy velocity fields are another important probe of the large
scale distribution of  matter.  The \citet{willick/strauss:1998} analysis of
the Mark III velocity fields and the IRAS redshift survey yields 
$\beta^{IRAS}=0.50 \pm 0.04$. 
IRAS galaxies are less clustered than optically selected
galaxies; \citet{fisher/etal:1994} find $\sigma_8^{IRAS} = 0.69 \pm 0.04$
implying $\sigma_8^{mass} \Omega_m^{0.6} = 0.345 \pm 0.05$, consistent
with our $\Lambda$CDM  model value of 
\ensuremath{0.44 \pm 0.10}.

\subsubsection{Cluster Number Counts}

Our best fit to the \map data is $\sigma_8 \Omega_m^{0.5} =0.48\pm 0.12$. 
\citet{bahcall/etal:2002} recent study of the mass function of 300 clusters at redshifts   $0.1 < z < 0.2$ in the early SDSS data release  yields $\sigma_8 \Omega_m^{0.5} = 0.33 \pm 0.03$. This difference may reflect
the sensitivity of the cluster measurements to the conversion of cluster richness to mass.
Observations of the mass function of high redshift clusters break the
degeneracy between $\sigma_8$ and $\Omega_m$.   The
recent \citet{bahcall/bode:2002} analysis of the
abundance of massive clusters at $z = 0.5 - 0.8$ yields $\sigma_8 = 0.95 \pm 0.1$ for $\Omega_m = 0.25$.
Other cluster analysis yield different values:
\citet{borgani/etal:2001} best fit values for
a large sample of X-ray clusters are $\sigma_8 = 0.66_{-0.05}^{+0.05}$
and $\Omega_m = 0.35^{+0.13}_{-0.10}$.  On the other hand,
\citet{reiprich/bohringer:2002} find very different values:
$\sigma_8 = 0.96^{+0.15}_{-0.12}$ and $\Omega_m = 0.12^{+0.06}_{-0.04}$.
\citet{pierpaoli/etal:2002} discuss the wide range of
values that different X-ray analyses find for $\sigma_8$.  With the larger REFLEX
sample, \citet{schuecker/etal:2003} find $\sigma_8 = 0.711_{-0.031}^{+0.039}\ _{-0.162}^{+0.120}$
and $\Omega_m = 0.341^{+0.031}_{-0.029}\ _{-0.071}^{+0.087}$, where the second set of
errors include the systematic uncertainties. The
best fit \map values lie in the middle of the relevant range.

Measurements of the contribution to the CMB power spectrum  on small
scales
from the Sunyaev-Zel'dovich effect also probe the number density of 
high redshift clusters.    
The recent CBI detection of excess fluctuations \citep{mason/etal:2001,bond/etal:2002} 
at $\ell >1500$ implies 
$\sigma_8 =  1.04 \pm 0.12$ \citep{komatsu/seljak:2002}, if the signal
is due to the Sunyaev-Zel'dovich effect.

\subsection{Baryon Abundance}

Both the amplitude of the acoustic peaks in the CMB spectrum 
\citep{bond/efstathiou:1984} and the primordial  abundance of  Deuterium 
\citep{boesgaard/steigman:1985} are sensitive functions of the cosmological
baryon density. Since the height and position of the acoustic peaks depend 
upon the properties of the cosmic plasma 372,000 years after
the Big Bang and the Deuterium 
abundance depends on physics only three minutes
after the Big Bang, comparing the baryon density 
constraints inferred from these two different probes provides an important test
of the Big Bang model. 
The best fit baryon abundance based on \map data only for the
PL LCDM model,
\ensuremath{\Omega_bh^2 = 0.024 \pm 0.001}, implies a baryon/photon ratio
of \ensuremath{\eta = (6.5^{+ 0.4}_{- 0.3}) \times 10^{-10} \mbox{ }}.
For this abundance, standard big bang nucleosynthesis 
\citep{burles/nollett/turner:2001} implies a primordial Deuterium 
abundance relative to Hydrogen: 
[D]/[H] $=  2.37^{+0.19}_{-0.21}\times 10^{-5}$. 
As it will be clear from \S 5  and  6,  the best fit $\Omega_b h^2$
value for our fits is relatively 
insensitive to cosmological model and data set combination as it depends primarily on the ratio of the first to second peak heights \citep{page/etal:2003c}.  
For the running spectral index model discussed in \S \ref{sec:running},
the best fit baryon abundance, \ensuremath{\Omega_bh^2 = 0.0224 \pm 0.0009},
implies a primordial [D]/[H] $=2.62^{+0.18}_{-0.20}\times 10^{-5}$.

\begin{deluxetable}{lll}
\tablecaption{Measured ratio of Deuterium to Hydrogen
\label{table:baryon}}
\tablehead{\colhead{Quasar} &  \colhead{[D]/[H]}
& \colhead{Reference}}
\startdata
Q0130-403 & $<6.8 \times 10^{-5}$  & \citet{kirkman/etal:2000} \\
PKS 1937-1009 & $3.25 \pm 0.3\times 10^{-5}$ & \citet{burles/tytler:1998a} \\
Q1009+299 & $4.0 \pm 0.65\times 10^{-5}$ & \citet{burles/tytler:1998b} \\
HS0105+1619 & $2.5 \pm 0.25\times 10^{-5}$ & \citet{omeara/etal:2001}\\
Q2206-199 & $1.65 \pm 0.35\times 10^{-5}$ & \citet{pettini/bowen:2001}\\
Q0347-383 & $3.75 \pm 0.25\times 10^{-5}$ & \citet{levshakov/etal:2003}  \\
Q1234+3047 & $2.42_{-0.25}^{+0.35} \times 10^{-5}$  & 
        \citet{kirkman/etal:2003} \\
\enddata
\end{deluxetable}

How does the primordial Deuterium abundance inferred from CMB compare with  
that  observed from the ISM?
 Galactic chemical evolution  destroys Deuterium because the 
Deuterium nucleus is relatively fragile and is easily destroyed in stars. 
Thus,  measurements of the Deuterium abundance within the Galaxy are usually treated as 
lower limits on the primordial abundance \citep{epstein/lattimer/schramm:1976}.
Local measurements  of D and H absorption find [D/H] abundance near 
$1.5 \times 10^{-5}$, while more distant measurements by IMAP and FUSE  
find significant variation in Deuterium abundances suggesting a complex 
Galactic chemical history \citep{jenkins/etal:1999, sonneborn/etal:2000, moos/etal:2002}.

Observations of Lyman $\alpha$ clouds reduce the need to correct the Deuterium
abundance for stellar processing as these systems have low (but non-zero) 
metal abundances.  These observations require identifying gas systems that 
do not have serious interference from the Lyman $\alpha$ forest. 
The \citet{kirkman/etal:2003} 
analysis of QSO HS 243+3057 yields a D/H ratio of 
$2.42_{-0.25}^{+0.35} \times 10^{-5}$.  They combine this
measurement  with four other D/H measurements 
(Q0130-4021: D/H$ < 6.8 \times 10^{-5}$, 
Q1009+2956: $3.98 \pm 0.70 \times 10^{-5}$,
PKS 1937-1009: $3.25 \pm 0.28 \times 10^{-5}$, and
QSO HS0105+1619: $2.5 \pm 0.25 \times 10^{-5}$), to obtain
their current best D/H ratio: $2.78 ^{+0.44}_{-0.38}\times 10^{-5}$ implying
$\Omega_b h^2 = 0.0214\pm 0.0020$. 
\citet{dodorico/dessauges-zavadsky/molaro:2001}  find $2.24\pm 0.67 \times 10^{-5}$ from their observations of Q0347-3819 (although  a reanalysis of the system by \citet{levshakov/etal:2003} finds
a higher D/H value: $3.75 \pm 0.25$.
\citet{pettini/bowen:2001} report a D/H abundance of  
$1.65 \pm 0.35 \times 10^{-5}$ from STIS measurements of QSO 2206-199, 
a low metallicity ($Z\sim 1/200$) Damped Lyman $\alpha$ system.  
The \map value lies between the \cite{pettini/bowen:2001}
estimate from DLAs, $\Omega_b h^2 = 0.025 \pm 0.001$,
 and the \cite{kirkman/etal:2003} estimate of
$\Omega_b h^2 = 0.0214 \pm 0.0020$ The remarkable agreement between the baryon
density inferred from  D/H values and our measurements is
an important triumph for the basic Big Bang model.

\subsection{Cosmic Ages}

\begin{deluxetable}{ll}
\tablecaption{
\label{tab:ages} Cosmic Age}
\tablehead{\colhead{Method} & \colhead{Age}}
\startdata
\map data ($\Lambda$CDM) & \ensuremath{13.4 \pm 0.3 \mbox{ Gyr}}\\
\mapext + LSS & \ensuremath{13.7 \pm 0.2 \mbox{Gyr}} \\
Globular Cluster Ages & $> 11- 16$ Gyr \\
White Dwarf & $> 12.7 \pm 0.7$ Gyr \\
OGLEGC-17 & $>10.4 - 12.8$ Gyr \\
Radioactive dating & $> 9.5 - 20$ Gyr \\
\enddata
\end{deluxetable}

The age of the Universe based on the best fit to \map data only,
\ensuremath{t_0 = 13.4 \pm 0.3 \mbox{ Gyr}}.
However, the addition of other data sets (see \S \ref{sec:combine}) implies
a lower matter density and a slightly larger age.  The best fit
age for the power law model based on a combination of \map, \tdf and
\lya forest data is \ensuremath{t_0 = 13.6 \pm 0.2 \mbox{Gyr}}.  The
best fit age for the same data set for the running index model of
\S \ref{sec:running} is \ensuremath{t_0 = 13.7 \pm 0.2 \mbox{Gyr}}.
(See \citet{hu/etal:2001} and \citet{knox/christensen/skordis:2001}
for discussions of using CMB data to determine cosmological ages.)

A lower limit to the age of the universe can independently be  obtained from dating the
oldest stellar populations. This has been done traditionally by dating
the oldest stars in the Milky Way  (see e.g., \citet{chaboyer:1998, jimenez:1999}).
For this program, globular clusters are an excellent laboratory for 
constraining the age of the universe: each cluster has a chemically 
homogeneous population of stars all born nearly simultaneously.
The main uncertainty in the age determination comes from the poorly known 
distance \citep{chaboyer:1995}. 
Well-understood stellar populations are useful tools for constraining cluster 
distances: \citet{renzini/etal:1996} used the white dwarf sequence to obtain an 
age of $14.5 \pm 1.5$ Gyr for NGC 6752. \citet{jimenez/etal:1996}.
using a distance-independent method determined the age of the oldest globular
clusters to be $13.5 \pm 2$ Gyr. Using the luminosity function method, 
\citet{jimenez/padoan:1998} 
found an age of $12.5\pm 1.0$ Gyr for M55. This method gives a joint constraint on the distance and the age of the globular cluster.
Other groups find consistent ages:  \citet{gratton/etal:1997} estimate 
an age of 11.8$^{+2.1}_{-2.5}$ Gyr for the oldest Galactic globulars; \citet{vandenBerg/etal:2002} estimates an age of $\sim 13.5$ Gyr for M92.  \citet{chaboyer/krauss:2003} review the globular cluster analysis and
quote a best fit age of 13.4 Gyr.

Observations of eclipsing double line spectroscopic binaries 
enable globular cluster age determinations that avoid the considerable
 uncertainty associated with
the globular cluster distance scale \citep{paczynski:1997}. 
\citet{thompson/etal:2001} were able to obtain a high precision 
mass estimate for the detached double line spectroscopic binary, 
OGLEGC-17 in $\omega-$Cen. Using the age/turnoff mass relationship,  
the \citet{kaluzny/etal:2002} analysis of this system yielded an age for
this binary of $11.8 \pm 0.6$ Gyr.  \citet{chaboyer/krauss:2002} 
re-analysis of the age/turnoff mass relationship for this system yields a 
similar age estimate: $11.1 \pm 0.67$ Gyr.
The \map determination of the age of the universe implies that globular clusters form within 2 Gyr after the Big Bang, a reasonable
estimate that is consistent with structure formation in the $\Lambda$CDM cosmology.
White dwarf dating provides an alternative approach to the traditional studies
of the main sequence turn-off.  \citet{richer/etal:2002} and 
\citet{hansen/etal:2002} find an age for the globular cluster M4 
of $12.7 \pm 0.7$ Gyrs (2 $\sigma$ errors, $\pm 0.35$ at the 1 $\sigma$ level 
assuming Gaussian errors) using the white dwarfs cooling sequence method. 
These results, which yield an age close to the cosmological age,
are potentially very useful: further tests of the assumptions
of the white dwarf age dating method will clarify its systematic
uncertainties.

Observations of nearby halo stars enable astronomers to obtain spectra of various radio-isotopes.
By measuring isotopic ratios, they infer stellar ages that are independent of much of the physics that determines main sequence turn-off
(see \citet{thielemann/etal:2002} for a recent review).
These studies yield stellar ages consistent with both the globular cluster ages and the ages in our best fit models.
\citet{clayton:1988} using a range of chemical evolution
models for the Galaxy finds ages between 12 - 20 Gyr. 
\citet{schatz/etal:2002}
study  Thorium and Uranium in CS 31082-001 and estimate
an age of $15.5 \pm 3.2$ Gyr for the r-process elements in the star. 
Other groups find similar estimates: 
the \citet{cayrel/etal:2001}
analysis of  U-238 in the old halo star CS
31082-001 yields an age of $12.5 \pm 3$ Gyr, while 
\citet{hill/etal:2002} 
find an age of $14.0 \pm 2.4$
Gyr.  Studies of other old halo stars
yield similar estimates:  \citet{cowan/etal:1999}
two stars CS 22892-052 and HD115444 find $15.6 \pm 4.6$ Gyr. 

Table \ref{tab:ages} summarizes the lower limits on the age
of the universe from various astronomical measurements.
While the errors on these measurements remain too large to effectively constrain parameters, they provide an important consistency check on our basic cosmological model.

\subsection{Large Scale Structure}

The large scale structure observations 
and the \lya\ forest data complement the CMB measurements
by measuring similar physical scales at very different epochs. The \map angular
power spectrum has the smallest uncertainties near $\ell \sim 300$, which
correspond to wavenumbers $k \sim 0.02$ Mpc$^{-1}$.  With the ACBAR results,
our CMB data set extends to $\ell \sim 1800$, corresponding to 
$k \sim 0.1~{\rm Mpc^{-1}}$.
If we assume that gravity is the primary force determining the large-scale
distribution of matter and that galaxies trace mass at least on large scales, 
then we can directly compare our best fit $\Lambda$ CDM  model (with
parameters fit to the \map data) to
observations of large scale distribution of galaxies.
There are currently
two major ongoing large scale structure surveys: the Anglo-Australian Telescope two degree field  Galaxy Redshift Survey (2dFGRS) \citep{colless/etal:2001}, and the Sloan Digital Sky Survey\footnote{www.sdss.org} (SDSS).
Large scale structure data sets are a powerful tool for breaking
many of the parameter degeneracies associated with CMB data.  
In \S 5, we make extensive use of the 2dFGRS  data set.

\begin{figure}
\figurenum{6}
\plottwo{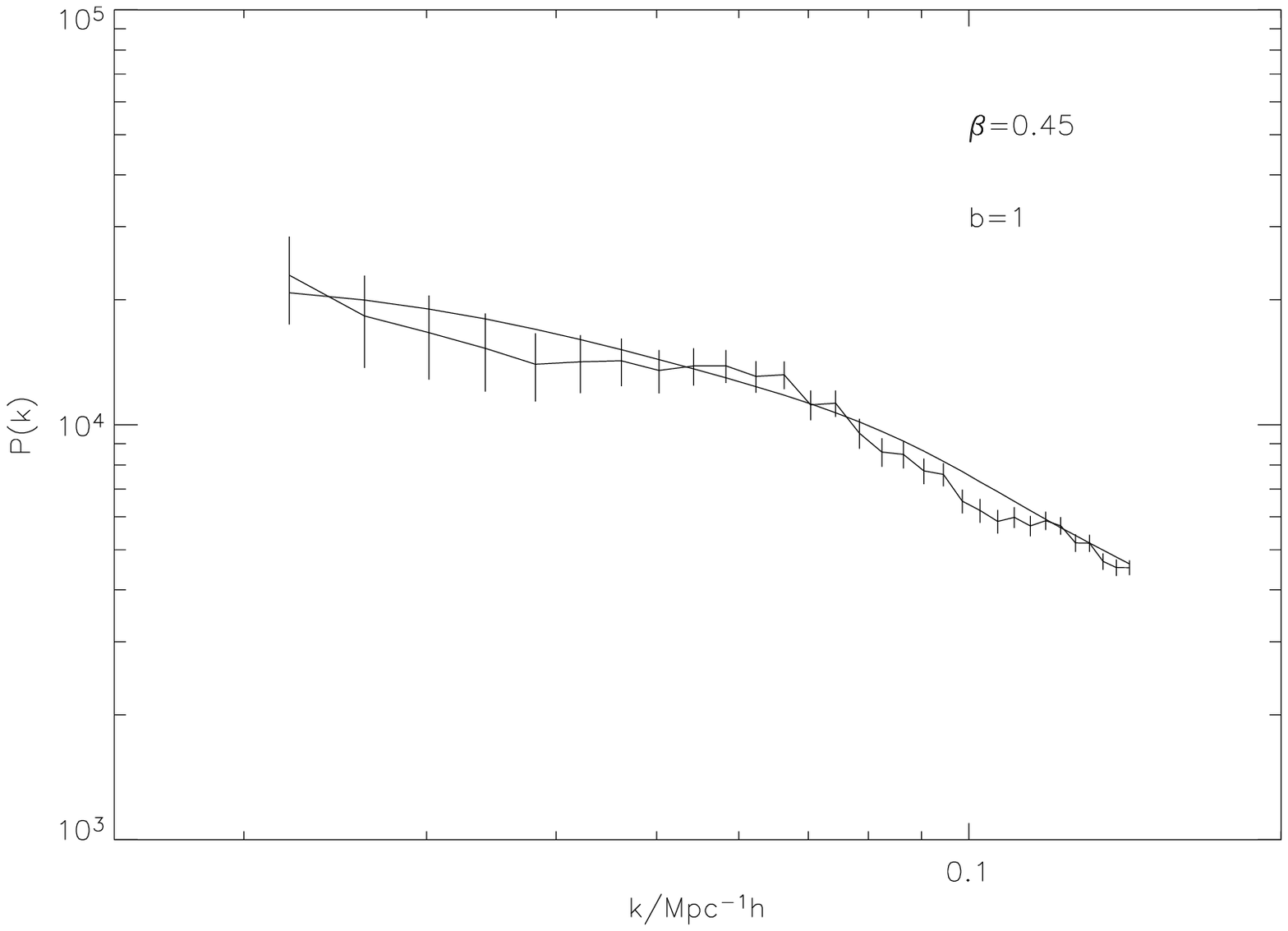}{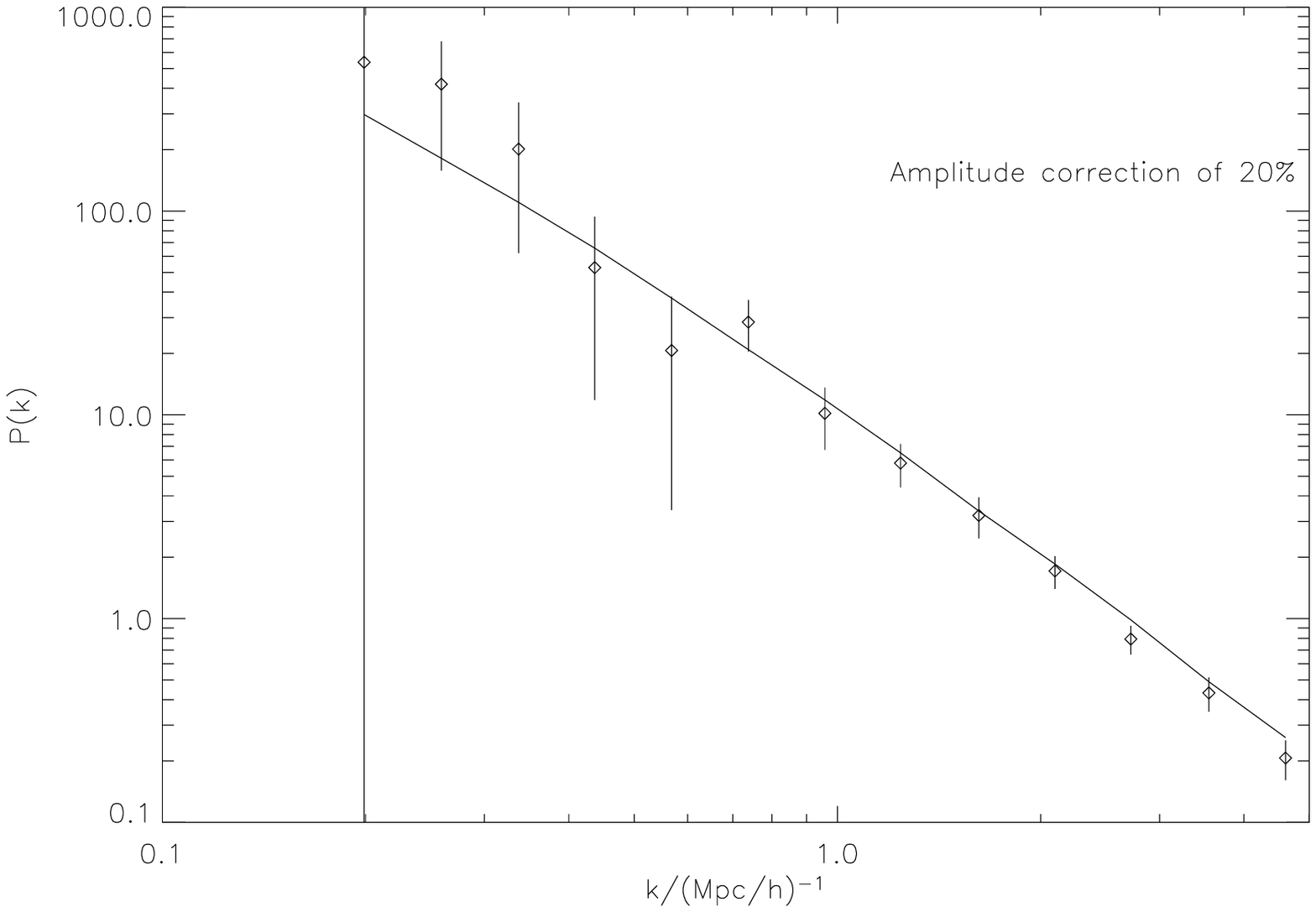}
\caption{(Left) This figure compares the best fit $\Lambda$CDM model of \S 3
based on {\sl WMAP} data only to the
2dFGRS Power Spectrum\citep{percival/etal:2001}.
The bias parameter for the best fit Power Law $\Lambda$CDM model is 1.0
corresponding to a best fit value of $\beta = 0.45$.
(Right) This figure compares the best fit Power Law $\Lambda$CDM model of \S 3
to the power spectrum at $z = 3$ inferred from the Lyman $\alpha$
forest data.  The data points have been scaled downwards by 20\%,
which is consistent with the 1 $\sigma$ calibration uncertainty
\citep{croft/etal:2002}.
\label{fig:2df}}
\end{figure}

Figure \ref{fig:2df} shows that the $\Lambda$CDM model obtained from the \map data alone is an acceptable fit to  the \tdf\ power spectrum.
The best fit has $\beta = 0.45$ consistent with \citet{peacock/etal:2001} 
measured value of  $\beta = 0.43 \pm 0.07$. 

The Lyman $\alpha$ forest observations are an important complement to CMB observations since they probe the linear  matter power spectrum at $z = 2 -3$
\citep{croft/etal:1998, croft/etal:2002}. These observations
are sensitive to  small length scales, inaccessible to
CMB experiments. 
Unfortunately, the relationship between the measured
flux power spectrum and the linear power spectrum is complex \citep{gnedin/hamilton:2002,croft/etal:2002}
and needs to be calibrated by numerical simulations.  In \citet{verde/etal:2003}, we describe
our methodology for incorporating the Lyman $\alpha$ forest data into
our likelihood approach.  
Figure \ref{fig:2df} compares the predicted power spectra for
the best fit $\Lambda$CDM model to the linear power spectra inferred by \citet{gnedin/hamilton:2002} and by \citet{croft/etal:2002}.

\subsection{Supernova Data}

Over the past decade, Type Ia supernovae have emerged as 
important cosmological probes.  Once supernova light curves have
been corrected using the correlation between decline rate and luminosity
\citep{phillips:1993, riess/press/kirshner:1995}
they appear to be remarkably good standard candles.
Systematic studies by the supernova cosmology project 
\citep{perlmutter/etal:1999} and by the high $z$ supernova search 
team \citep{riess/etal:1998}
provide evidence for an accelerating universe.  The combination
of the large scale structure, CMB and supernova data provide strong
evidence for a flat universe dominated by a cosmological constant
\citep{bahcall/etal:1999}.
Since the supernova data probes the luminosity distance versus redshift relationship
at moderate redshift $z < 2$ and the CMB data probes the angular diameter distance
relationship to high redshift ($z \sim 1089$) , the two data sets are complementary. The supernova constraint on cosmological parameters  are consistent with the $\Lambda$CDM  \map model.
 As we will see in the discussion of non-flat models and quintessence models,
 the SNIa likelihood surface in the $\Omega_m-\Omega_\Lambda$  and in the $\Omega_m-w$ planes  provides useful additional constraints on cosmological parameters.

\subsection{Reionization \& Small Scale Power}
\label{sec:reionization}

The \map  detection of reionization \citep{kogut/etal:2003} implies the existence
of an early generation of stars able to reionize the Universe at $z\sim 20$.
Is this early star formation
compatible with our best fit  $\Lambda$CDM cosmological model?
We can evaluate this effect by first computing the fraction
of collapsed objects, $f_{DM}$, at a given redshift:
\begin{equation}
f_{DM}(z)=\frac{1}{\rho_0}\int_{M_{min}}^{\infty}\Phi(M,z)MdM,
\label{eq:massfn}
\end{equation}
where $\Phi(M,z)$ is the \citet{sheth/tormen:1999} mass function.
The first stars correspond to extremely rare fluctuations of the overdensity field: Eq. (\ref{eq:massfn}) is very sensitive to the tail of the mass function.  Thus the  very small change in the
minimum mass needed for star formation results in a significant change in the fraction of collapsed objects.
The minimum halo mass for star formation, $M_{min}$, 
is controversial and depends on
whether molecular hydrogen (H$_2$) is available as a coolant.
If the gas temperature is fixed to the CMB temperature, then
the Jean Mass, $M^j = 10^6 M_\odot$.
If molecular hydrogen is available, then the Jeans mass before reionization 
is $M^{j'}\sim 2.2 \times 10^3 [\omega_b/h(\omega_m)]^{1.5}(1+z)/10$
for $z < 150$
\citep{venkatesan/giroux/shull:2001}. At $z > 150$, the electrons
are thermally coupled to the CMB photons.
However, as \citet{haiman/rees/loeb:1997} point out, a small
UV background generated by the first sources will dissociate $H_2$, thus making
the minimum mass much larger than the Jeans mass.
They suggest using a  minimum mass that is much higher: 
$M^{HRL}_{min}(z)= 10^8(1+z)/10)^{-3/2}$.
On the other hand if the first stars generated a significant flux
of  X-rays \citep{oh:2001} then this would have promoted molecular hydrogen formation
\citep{haiman/abel/rees:2000,venkatesan/giroux/shull:2001,cen:2002}. Thus lowering the minimum mass  back to $M^{j}$. 

Following \citet{tegmark/silk:1995} we estimate the rate of reionization by multiplying the
collapse factor by an efficiency factor.
A fraction of baryons in the universe, $f_b$, falls into the non-linear
structures. We assume $f_b=f_{DM}$ (i.e., constant baryon/dark matter ratio).
A certain fraction of these baryons form stars or quasars, $f_{burn}$, which
emit UV radiation with some efficiency, $f_{UV}$. Some of this radiation escapes
into the intergalactic medium
photoionizing it; however, the net number of ionizations per UV photons,
$f_{ion}$,  is
expected to be less than unity (due to cooling and recombinations).
Finally the intergalactic medium might be  clumpy, making the photoionization
process less efficient. This effect is counted for by the clumping factor $C_{clump}$.
Thus  in this approximation the ionization fraction is given
by:  $x_e = 3.8\times 10^5 f_{net}f_b$ where
$f_{net}=f_{burn}f_{UV}f_{esc}f_{ion}/C_{clump}$. 
The factor  $3.8\times 10^5$ arises because $ 7.3\times 10^{-3}$ 
of the rest mass is released in the burning of hydrogen
to helium and we assume the primordial helium mass fraction to be 24\%. We
assume
$f_{burn}\lap 25\%$, $f_{esc}\lap 50\%$, $f_{UV}\lap 50\%$, $f_{ion}\lap
90\%$, and $1\lap C_{clump}\lap 100$, thus $f_{net}\lap 5.6\times 10^{-3}$.

\begin{figure}
\figurenum{7}
\plotone{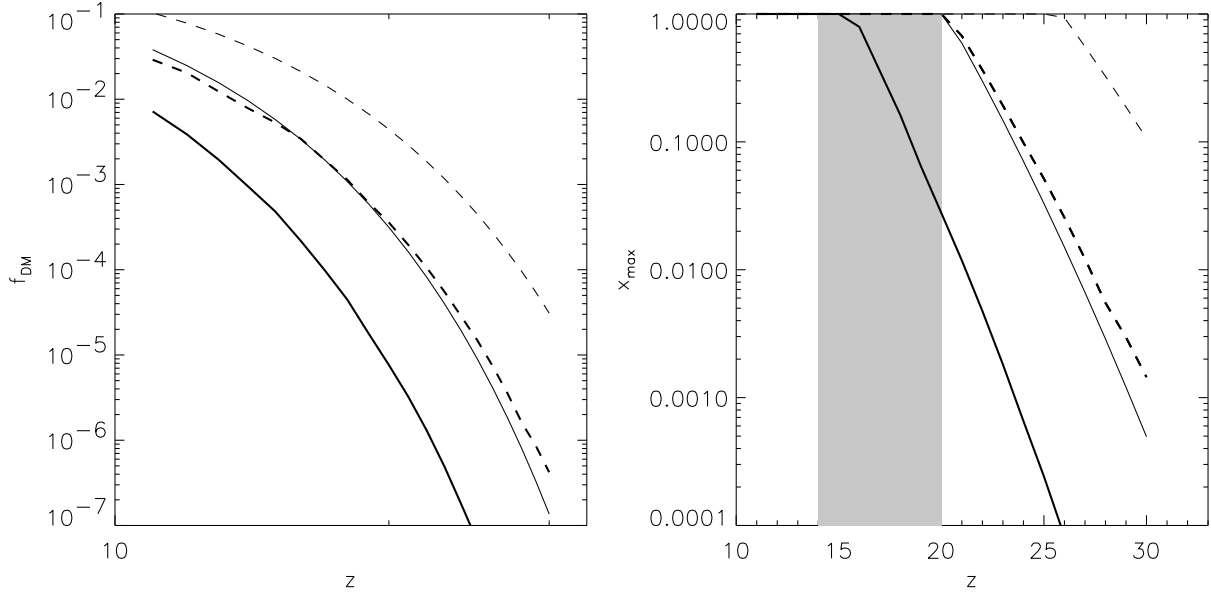}
\caption{
\label{fig:reion}
(Left panel)
This figure shows the  fraction of mass in bound objects
as a function of redshift.  The black lines show
the mass in collapsed objects with mass greater than $M^{HRL}(z)$, the
effective Jeans mass in the absence of $H_2$ cooling for our
best fit PL $\Lambda$CDM model (thin lines are for the fit to {\sl WMAP} only and
thick lines are  for the fit to all data sets). The heavy line uses
the best fit parameters based on all data (which has a lower $\sigma_8$)
and the light line uses the best fit parameters based
on fitting to the \map data only.
The dashed  lines show the mass in collapsed objects
with masses greater than the Jeans mass assuming that 
the minimum mass is $10^6 M_\odot$.  More objects form
if the minimum mass is lower.
(Right Panel) This figure shows the  ionization fraction
as a function of redshift.  The solid line shows
ionization fraction for the best fit PL $\Lambda$CDM model
if we assume that $H_2$ cooling is suppressed by photo-destruction 
of $H_2$.  This figure suggests that $H_2$ cooling
may be necessary for enough objects to form
early enough to be consistent with the \map detection.
The heavy line is for the best fit parameters for all data sets
and the light line is for the best fit parameters for the
\map only fit.  The dashed lines assume that the objects
with masses greater than $10^6 M_\odot$ can form stars.
The gray band shows the 68\% likelihood region for
$z_{r}$ based on the assumption of instantaneous complete
reionization \citep{kogut/etal:2003}.}
\end{figure}

Figure \ref{fig:reion} shows the  fraction of collapsed objects and the maximum ionization fraction as a function
of redshift for  our best fit \map $\Lambda$CDM model. The solid lines correspond to $M_{min}=M^{HRL}_{min}(z)$ while the dashed lines correspond to $M_{min}=M^{j}$.   The \map detection of reionization at high redshift suggests
that $H_2$ cooling likely played an important role in early star formation.

Because early reionization requires the existence
of small scale fluctuations,
the \map TE detection has important implications for our understanding of 
the nature of the dark matter.  
\citet{barkana/haiman/ostriker:2001} note that the detection of reionization
at $z > 10$ rules out warm dark matter as a viable candidate for the
missing mass as structure forms
very late in these models.
Warm dark matter can not cluster on scales smaller than
the dark matter Jeans' mass.  Thus, this limit applies regardless of whether the
minimum mass is $M^{HRL}$ or $M^{j}$.

\section{COMBINING DATA SETS}
\label{sec:combine}

In this section, we combine the \map data with other CMB
experiments that probe smaller angular scales (ACBAR and CBI)
\footnote{In the following sections, we refer to the combined \map, ACBAR and CBI
data sets as \mapext.}
and with astronomical measurements of the power spectrum (the \tdf\ and \lya\ forest).  
We begin by
exploring how including these data sets affects our best fit power law
$\Lambda$CDM model parameters (\S \ref{sec:plcdm}). The addition of data sets
that probe smaller scales systematically pulls down the amplitude
of the fluctuations in the best fit model.  This motivates our
exploration of an  extension
of the power law model, a model where the primordial power spectrum
of scalar density fluctuations
is fit by a running spectral index \citep{kosowsky/turner:1995}:
\begin{equation}
P(k) = P(k_0) \left(\frac{k}{k_0}\right)^
{n_s(k_0) + (1/2){dn_s/d\ln k} \ln(k/k_0)},
\end{equation}
where we fix the scalar spectral index and slope at $k_0 = 0.05$Mpc$^{-1}$.
Note that this definition of the running index matches
the definition used in \citet{hannestad/etal:2002} analysis
of running spectral index models
and differs by a factor of 2 from the \citet{kosowsky/turner:1995} definition.
As in the scale independent case, we define
\begin{equation}
n_s(k) = \frac{d\ln P}{d\ln k}.
\end{equation}
We explicitly assume that $d^2n_s/dlnk^2 = 0$, so that
\begin{equation}
n_s(k) = n_s(k_0) + \frac{d n_s}{d\ln k} \ln\left(\frac{k}{k_0}\right).
\end{equation}

In \S \ref{sec:running}, we show that the running spectral 
index model is a better
fit than the pure power law model 
to the combination of \map and other data sets.  
\citet{peiris/etal:2003} explores the 
implications of this running spectral index
for inflation.

\subsection{Power Law CDM Model}
\label{sec:plcdm}

\begin{deluxetable}{lllll}
\tablecaption{Best Fit Parameters: Power Law $\Lambda$ CDM
\label{tab:plcdm}}
\tablewidth{0pt}
\tablehead{
\colhead{} & \colhead{\map} &\colhead{\mapext}\footnotemark{a} &  
\colhead{\mapext+\tdf}  & \colhead{\mapext + \tdf + \lya}}
\startdata
$A$ &\ensuremath{0.9 \pm 0.1} &
\ensuremath{0.8 \pm 0.1}&
\ensuremath{0.8 \pm 0.1}&
\ensuremath{0.75^{+ 0.08}_{- 0.07}} \\
$n_s$ &\ensuremath{0.99 \pm 0.04} &
\ensuremath{0.97 \pm 0.03}&
\ensuremath{0.97 \pm 0.03}&
\ensuremath{0.96 \pm 0.02} \\
$\tau$ &\ensuremath{0.166^{+ 0.076}_{- 0.071}} &
\ensuremath{0.143^{+ 0.071}_{- 0.062}}&
\ensuremath{0.148^{+ 0.073}_{- 0.071}}&
\ensuremath{0.117^{+ 0.057}_{- 0.053}} \\
$h$ &\ensuremath{0.72 \pm 0.05} &
\ensuremath{0.73 \pm 0.05}&
\ensuremath{0.73 \pm 0.03}&
\ensuremath{0.72 \pm 0.03} \\
$\Omega_m h^2$ &\ensuremath{0.14 \pm 0.02} &
\ensuremath{0.13 \pm 0.01}&
\ensuremath{0.134 \pm 0.006}&
\ensuremath{0.133 \pm 0.006} \\
$\Omega_b h^2$ &\ensuremath{0.024 \pm 0.001} &
\ensuremath{0.023 \pm 0.001}&
\ensuremath{0.023 \pm 0.001}&
\ensuremath{0.0226 \pm 0.0008} \\
$\chi^2_{eff}/\nu$ &1429/1341  &1440/1352 &1468/1381 & \nodata\tablenotemark{b} \\
\enddata
\tablenotetext{a}{\map+CBI+ACBAR}
\tablenotetext{b}{Since the Lyman $\alpha$ data points are correlated, we
do not quote an effective $\chi^2$ for the combined likelihood
including Lyman $\alpha$ data (see \citet{verde/etal:2003}).}
\end{deluxetable}

The power law $\Lambda$CDM model is an acceptable fit to the \map data.  
While it
overpredicts the amplitude of fluctuations on large angular scales (see \S 6),
this deviation may be due to cosmic variance at these large scales.  
Intriguingly, it also overpredicts the amplitude of fluctuations on small
angular scales.  

Table (\ref{tab:plcdm}) shows the best fit parameters for the power 
law $\Lambda$CDM
model for different combination of data sets.  As we add more and more data
on smaller scales, the best fit value for the amplitude of fluctuations
at $k = 0.05$ Mpc$^{-1}$
gradually drops:  When we fit to the \map data alone, the best fit is 
\ensuremath{0.9 \pm 0.1}.  When we add the CBI, ACBAR and \tdf\ data,
the best fit value drops to \ensuremath{0.8 \pm 0.1}.
Adding the Lyman $\alpha$ data further reduces $A$ to
\ensuremath{0.75^{+ 0.08}_{- 0.07}}.
The best fit spectral index shows a similar trend: the addition of more and
more small scale data drives the best fit spectral index to also change
by nearly 1 $\sigma$ from its best fit value for \map data only:
\ensuremath{0.99 \pm 0.04} (\map only) to
\ensuremath{0.96 \pm 0.02} ({\sl WMAP}ext+\tdf+Ly$\alpha$).
When the addition of new data continuously pulls a model away
from its best fit value, this is often the signature of the model
requiring a new parameter.  

\subsection{Running Spectral Index $\Lambda$CDM Model}
\label{sec:running}

\begin{deluxetable}{lllll}
\tabletypesize{\footnotesize}
\tablecaption{Best Fit Parameters for the Running Spectral Index 
$\Lambda$CDM Model
\label{tab:run}}
\tablewidth{0pt}
\tablehead{
\colhead{} & \colhead{\map } &\colhead{\mapext } & 
\colhead{\mapext +\tdf }  & \colhead{\mapext + \tdf + \lya }}
\startdata
$A$ &\ensuremath{0.92 \pm 0.12} &
\ensuremath{0.9 \pm 0.1}&
\ensuremath{0.84 \pm 0.09}&
\ensuremath{0.83^{+ 0.09}_{- 0.08}} \\
$n_s$ &\ensuremath{0.93 ^{+0.07}_{-0.07}} &
\ensuremath{0.91 \pm 0.06}&
\ensuremath{0.93^{+ 0.04}_{- 0.05}}&
\ensuremath{0.93 \pm 0.03} \\
$dn_s/d\ln k$ &\ensuremath{-0.047 \pm 0.04} &
\ensuremath{-0.055 \pm 0.038}&
\ensuremath{-0.031^{+ 0.023}_{- 0.025}}&
$-0.031_{-0.017}^{+0.016}$ \\
$\tau$ &\ensuremath{0.20 \pm 0.07} &
\ensuremath{0.20 \pm 0.07}&
\ensuremath{0.17 \pm 0.06}&
\ensuremath{0.17 \pm 0.06} \\
$h$ &\ensuremath{0.70 \pm 0.05} &
\ensuremath{0.71 \pm 0.06}&
\ensuremath{0.71 \pm 0.04}&
\ensuremath{0.71^{+ 0.04}_{- 0.03}} \\
$\Omega_m h^2$ &\ensuremath{0.14 \pm 0.02} &
\ensuremath{0.14 \pm 0.01}&
\ensuremath{0.136 \pm 0.009}&
\ensuremath{0.135^{+ 0.008}_{- 0.009}} \\
$\Omega_b h^2$ &\ensuremath{0.023 \pm 0.002} &
\ensuremath{0.022 \pm 0.001}&
\ensuremath{0.022 \pm 0.001}&
\ensuremath{0.0224 \pm 0.0009} \\
$\chi^2_{eff}/\nu$ &  1431/1342  &1437/1350 &1465/1380 & *\tablenotemark{a}\\
\enddata
\tablenotetext{a}{Since the Lyman $\alpha$ data points are correlated, we
do not quote  $\chi^2_{eff}$ for the combined likelihood
including Lyman $\alpha$ data (see \citet{verde/etal:2003}).}
\end{deluxetable}

Inflationary models predict that the spectral index of fluctuations
should be a slowly varying function of scale.  \citet{peiris/etal:2003}
discusses the inflationary predictions and shows that a plausible set
of models predicts a detectable varying spectral index.
There are classes of inflationary models that predict
minimal tensor modes. This section explores this class of models.
In \S \ref{sec:tensors}, we explore a more general model that has
both a running spectral index and tensor modes.

Table {\ref{tab:run}} shows the best fit parameters for the running (RUN) spectral
index model as a function of data set.
Note that the best fit parameters for
these models barely change as we add new data sets; however, the error
bars shrink.  When we include all data sets, the best fit value of
the running of the spectral index is 
$-0.031_{-0.017}^{+0.016}$:
fewer than 5\% of the models have $dn_s/d\ln k > 0$.

\begin{figure}
\figurenum{9}
\plottwo{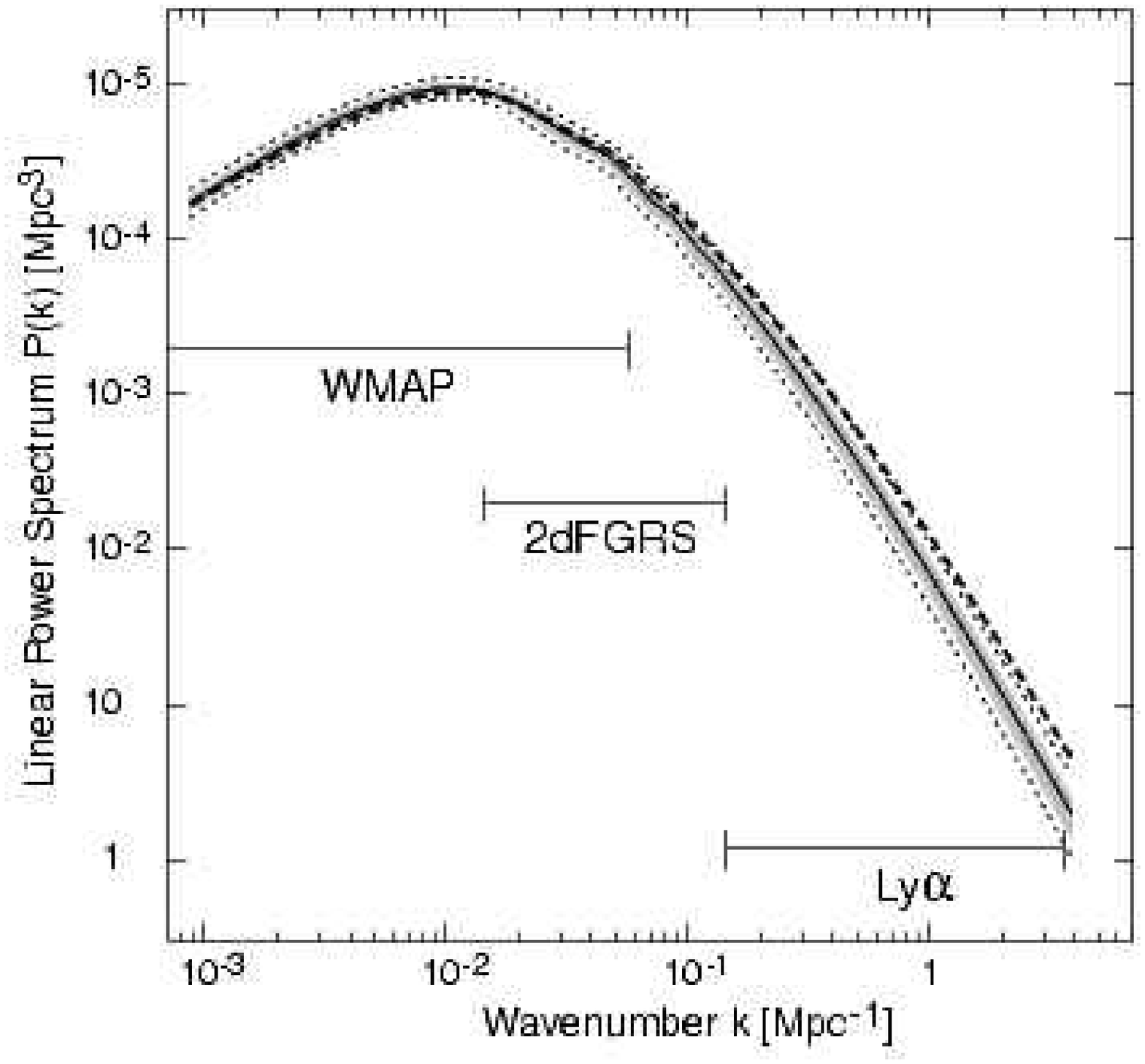}{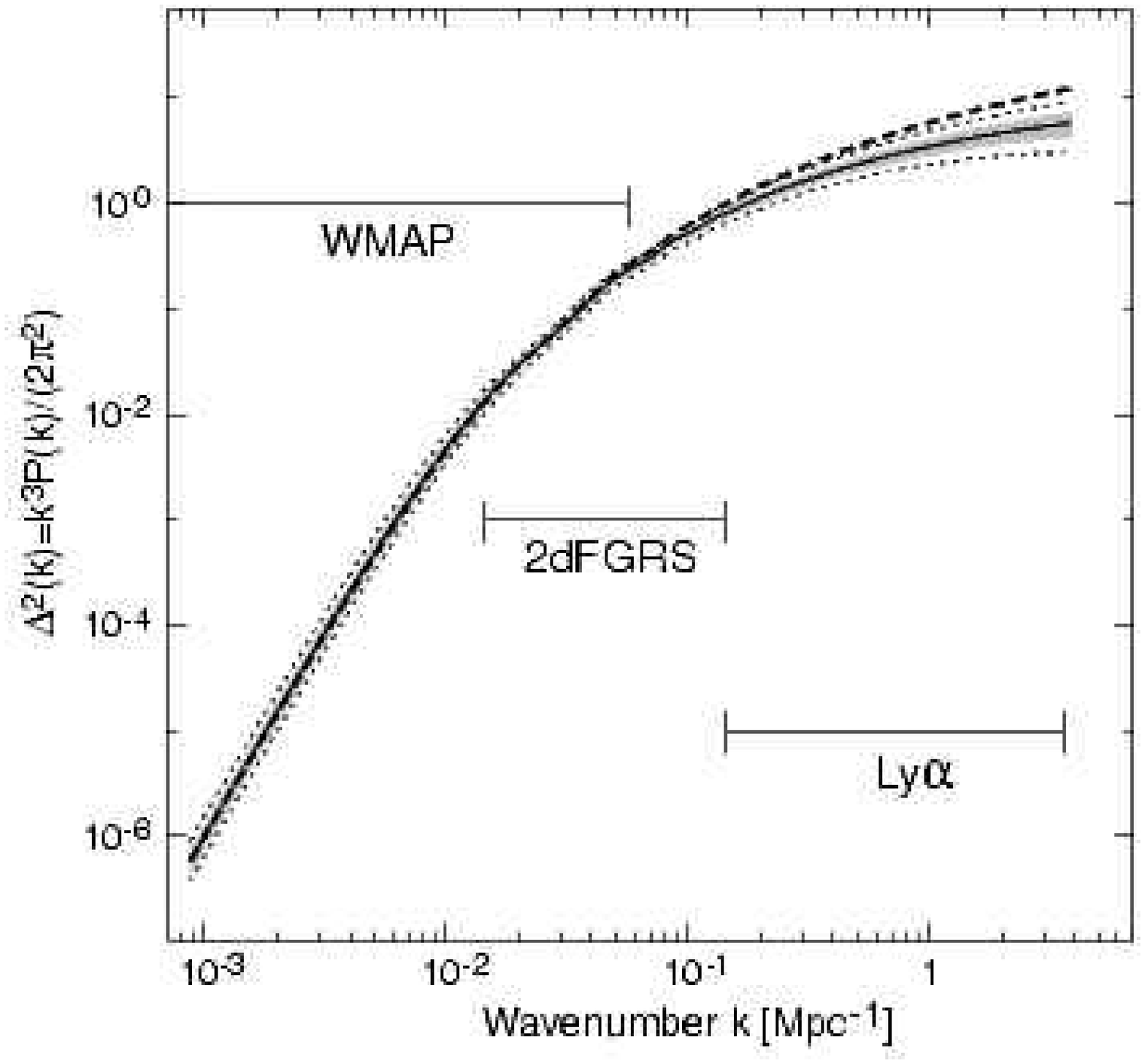}
\caption{
(Left)The shaded region in the figure shows the $1-\sigma$
contours for the  amplitude of the power spectrum as a function of scale
for the running spectral index model fit to all data sets.  The dotted
lines bracket the 2-$\sigma$ region for this model.  The dashed
line is the best fit power spectrum for the power law $\Lambda$CDM
model.
(Right)The shaded region in the figure shows the 1 -$\sigma$
contours for the  amplitude of the amplitude of mass
fluctuations, $\Delta^2(k) = (k^3/(2\pi^2) P(k)$, as a function of scale
for the running spectral index model fit to all data sets.  The dotted
lines bracket the 2-$\sigma$ region for this model.  The dashed
line is the best fit for the power law $\Lambda$CDM
model.
\label{fig:pk}}
\end{figure}

Figure \ref{fig:pk} shows the the power spectrum as a function of
scale. The figure shows the results of our Markov chain analysis
of the combination of \map, CBI, ACBAR, \tdf\ and \lya data.
At each wavenumber, we compute the range of values for the
power law index for all of the points in the Markov chain.  
The 68\% and 95\% contours at each $k$ value are shown
in Figure \ref{fig:pk} for the fit to the \mapext+\tdf\ + \lya 
data sets.  

Over the coming year, new data will significantly improve our ability
to measure (or constrain) this running spectral index.  When we complete
our analysis of the EE power spectrum, the \map data will place
stronger constraints on $\tau$.  Because of the $n_s-\tau$ degeneracy,
this implies a strong constraint on $n_s$ on large scales.  The SDSS collaboration will
soon release its galaxy spectrum and its measurements of the Lyman $\alpha$
forest.  These observations will significantly improve our measurements of $n_s$
on small scales.  \citet{peiris/etal:2003} shows that the detection of a running
spectral index and particularly the detection of a spectral index
that varies from $n_s > 1$ on large scales to $n_s < 1$ on small scales
would severely constrain inflationary models.

\begin{figure}
\figurenum{10}
\plotone{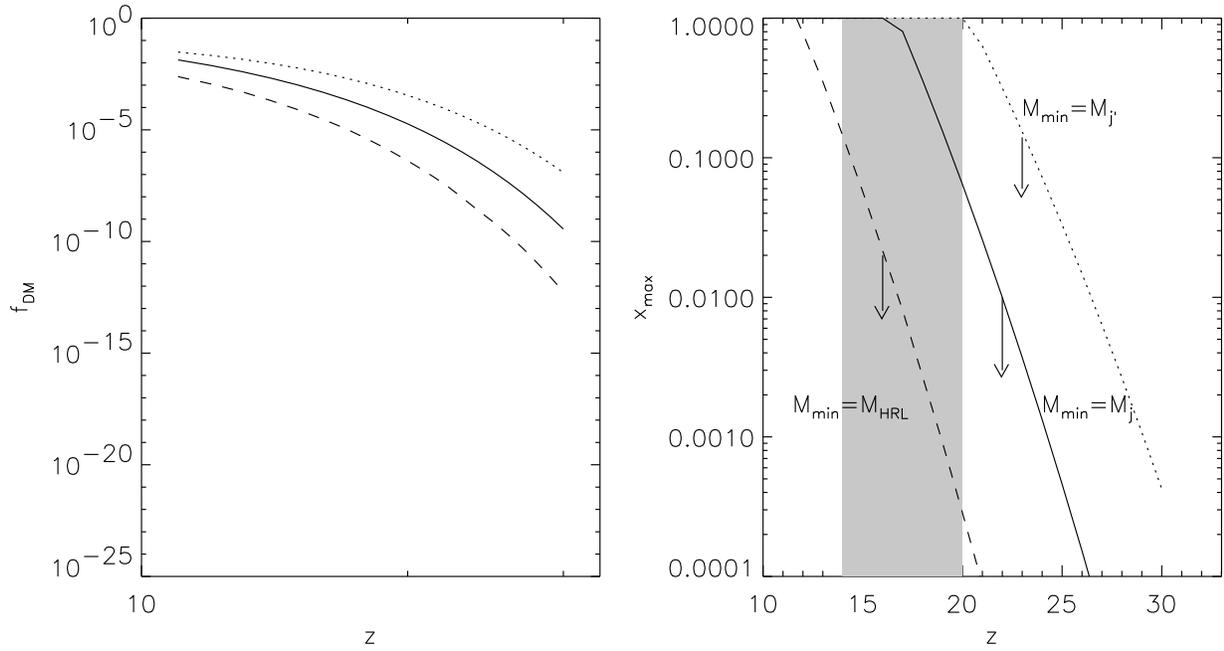}
\caption{
\label{fig:reiona}
(Left) This figure shows the fraction of the universe
in bound objects with mass greater than $M^{HRL}$ (dashed),
$M^j = 10^6 M_\odot$ (solid) and $M^{j'}$ (dotted) in a model
with a running spectral index.  The curves were computed
for the $1\sigma$ upper limit parameters for this model
(see Figure \ref{fig:pk}).  These should be viewed as
upper limits on the mass fraction in collapsed objects.
(Right) This figure shows the ionization fraction as
a function of redshift and is based on the assumptions
described in \S \ref{sec:reionization}.  As in the figure
on the left, we use the $1\sigma$ upper limit estimate
of the power spectrum so that we obtain "optimistic" estimates
of the reionization fraction.  
In the context of a running spectral index fit to the data,
the \map detection of reionization appears to require that 
$H_2$ cooling played an important role in early star formation.}
\end{figure}

The running spectral index model predicts  a significantly lower amplitude
of fluctuations on small scales than the standard $\Lambda$CDM model
(see figure \ref{fig:pk}).
This suppression of small scale power has several important astronomical
implications: (a) the reduction in small scale power makes it more
difficult to reionize the universe unless $H_2$ cooling enables 
mass dark halos to collapse and form galaxies (see \S \ref{sec:reionization}
and Figure \ref{fig:reiona});
(b) a reduction in
the small scale power reduces the amount of substructure within
galactic halos \citep{zentner/bullock:2002} (c)
since small objects form later, their dark
matter halos will be less concentrated
as there is a monotonic relationship between collapse time
and halo central concentration \citep{navarro/frenk/white:1997,
eke/navarro/steinmetz:2001,zentner/bullock:2002,wechsler/etal:2002,huffenberger/seljak:2003}.  The
reduction in the amount of substructure will also reduce angular momentum
transport between dark matter and baryons and will also reduce 
the rate of disk destruction through infall \citep{toth/ostriker:1992}.
We suspect that our proposed modification of the
primordial power spectrum will resolve many of the long-standing
problems of the CDM model on small scales (see \citet{moore:1994}
and \citet{spergel/steinhardt:2000} for discussions of the 
failings of the power law $\Lambda$ CDM model on galaxy  scales).

\section{BEYOND THE $\Lambda$CDM MODEL}

In this section, we consider various extensions to the $\Lambda$CDM model.
In \S \ref{sec:dark}, we consider dark energy models with a constant equation
of state.  In \S \ref{sec:nf}, we consider non-flat models.  In \S \ref{sec:mnu},
we consider models with a massive light neutrino.  In \S \ref{sec:tensors},
we include tensor modes.

In this section of the paper, we combine the \map data with external data sets
so that we can break degeneracies and 
obtain significant constraints on the various extensions
of our standard cosmological model.
\subsection{Dark Energy}
\label{sec:dark}
The properties of the dark energy, the dominant component in our universe today,
is a mystery.  The most popular alternative to the cosmological constant is
quintessence.  \citet{wetterich:1988}, \citet{ratra/peebles:1988} and 
\citet{peebles/ratra:1988} suggest that a rolling
scalar field could produce a time-variable dark energy
term, which leave a characteristic imprint on the CMB and on large scale structure  \citep{caldwell/dave/steinhardt:1998}.
In these quintessence models, the dark energy properties
are quantified by the equation of state of the dark energy:
$w = p/\rho$, where $p$ and $\rho$ are the pressure
and the density of the dark energy.  A cosmological constant
has an equation of state, $w = -1$.

Since the space of possible models is quite large, we only consider models
with a constant equation of state.  We now increase our model space
so that we have 7 parameters in the cosmological model
($A, n_s, h, \Omega_m, \Omega_b, \tau$ and $w$). We analyze
the data using two approaches: (a) we begin by restricting our analysis to $w > -1$ motivated by the
difficulties in constructing stable models with $w < -1$
\citep{carroll/hoffman/trodden:2003} and (b) relax this constraint and consider models
that violate the weak energy condition 
\citep{schuecker/etal:2003b}. Further
analysis is needed for models where $w$ and the quintessence
sound speed are a function of time \citep{dedeo/caldwell/steinhardt:2003}.
The addition of a new
parameter introduces a new degeneracy between $\Omega_m$, $h$,
and $w$ that can not be broken by CMB data alone \citep{huey/etal:1999,verde/etal:2003}:
models with the same values of $\Omega_m h^2$, $\Omega_b h^2$
and first peak position have nearly identical angular power spectra.

For example, a model with $\Omega_m = 0.47, w = -1/2$ and $h=0.57$ has
a nearly identical angular power spectrum to our $\Lambda$CDM  model. 
Note, however, that this Hubble Constant value differs by $2\sigma$  from 
the HST Key Project value and the predicted
shape of the power spectrum
is a poor fit to the \tdf\ observations.
This model is also a  worse fit to the supernova angular diameter distance relation.

We consider four different combinations of astronomical data
sets: (a) \mapext\ data  combined with the supernova observations;
(b) \mapext\ data combined with HST data;
(c) \mapext\ data combined with the \tdf\ large scale structure
data; 
(d) all data sets combined.

\begin{figure}
\figurenum{11}
\plotone{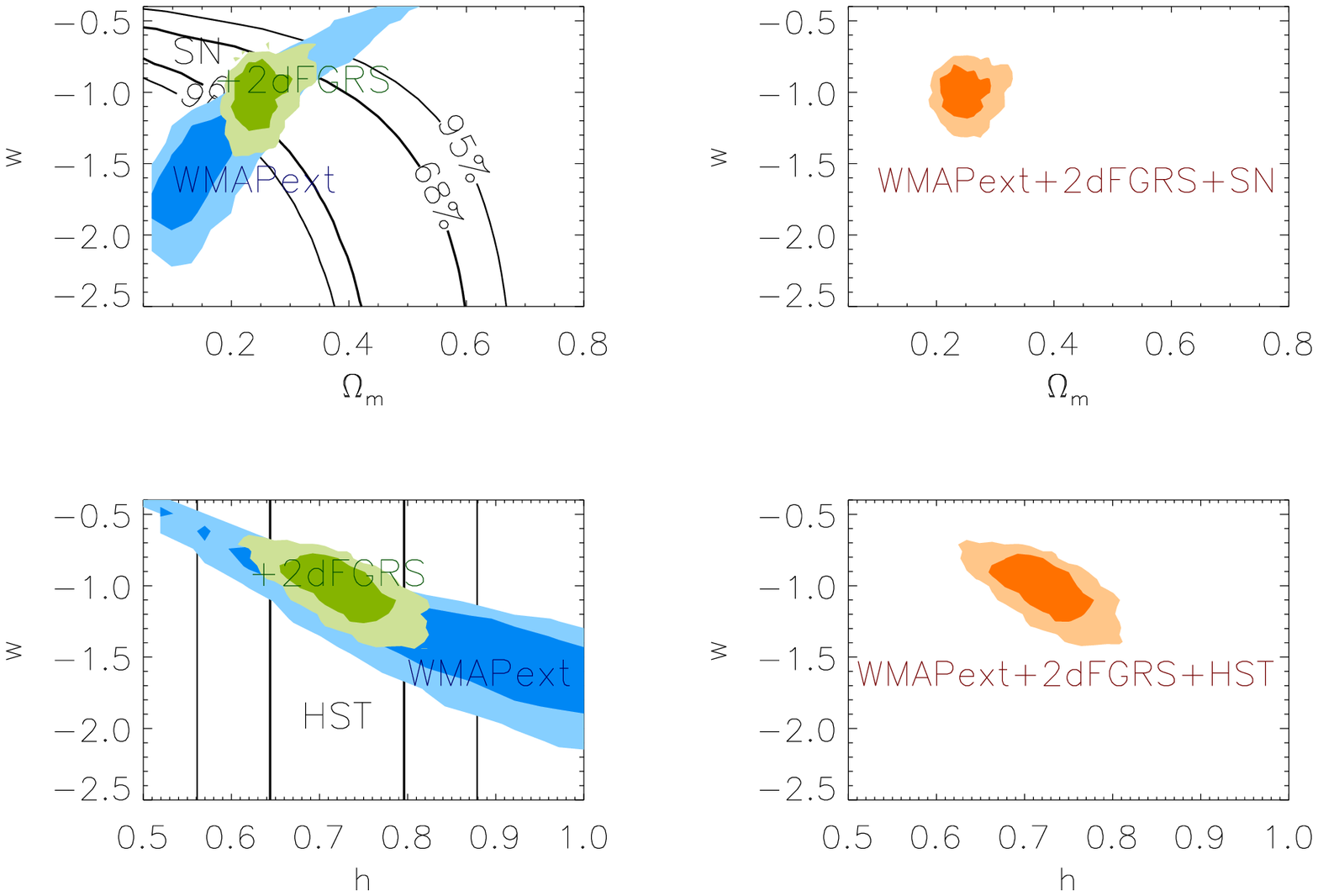}
\caption{Constraints on Dark Energy Properties.   The upper left panel 
shows the marginalized
maximum likelihood surface for the \mapext\  data alone
and for a combination of the \mapext\ + \tdf\ data sets.   The solid
lines in the figure show the 68\% and 95\% confidence ranges 
for the fit C supernova data from \citet{perlmutter/etal:1999}.
In the upper right panel, we
multiply  the supernova likelihood function by  the \mapext\ + \tdf\ likelihood
functions. The  lower left
panel shows the maximum likelihood surface for
$h$ and $w$ for the \mapext\ data alone and
 for the \mapext\ + \tdf\ data sets.  The solid lines in
the figures are the 68\% and 95\% confidence limits on $H_0$
from the HST Key Project, where we add the systematic and statistical
errors in quadrature.  In the lower right panel, we multiply the
likelihood function for the \mapext\ + \tdf\ data by the likelihood
surface for the HST data to determine the joint likelihood surface.
The dark areas in these plots are the 68\% likelihood regions and the light
areas are the 95\% likelihood regions.  
\label{fig:w}}
\end{figure}

\begin{figure}
\figurenum{12}
\plotone{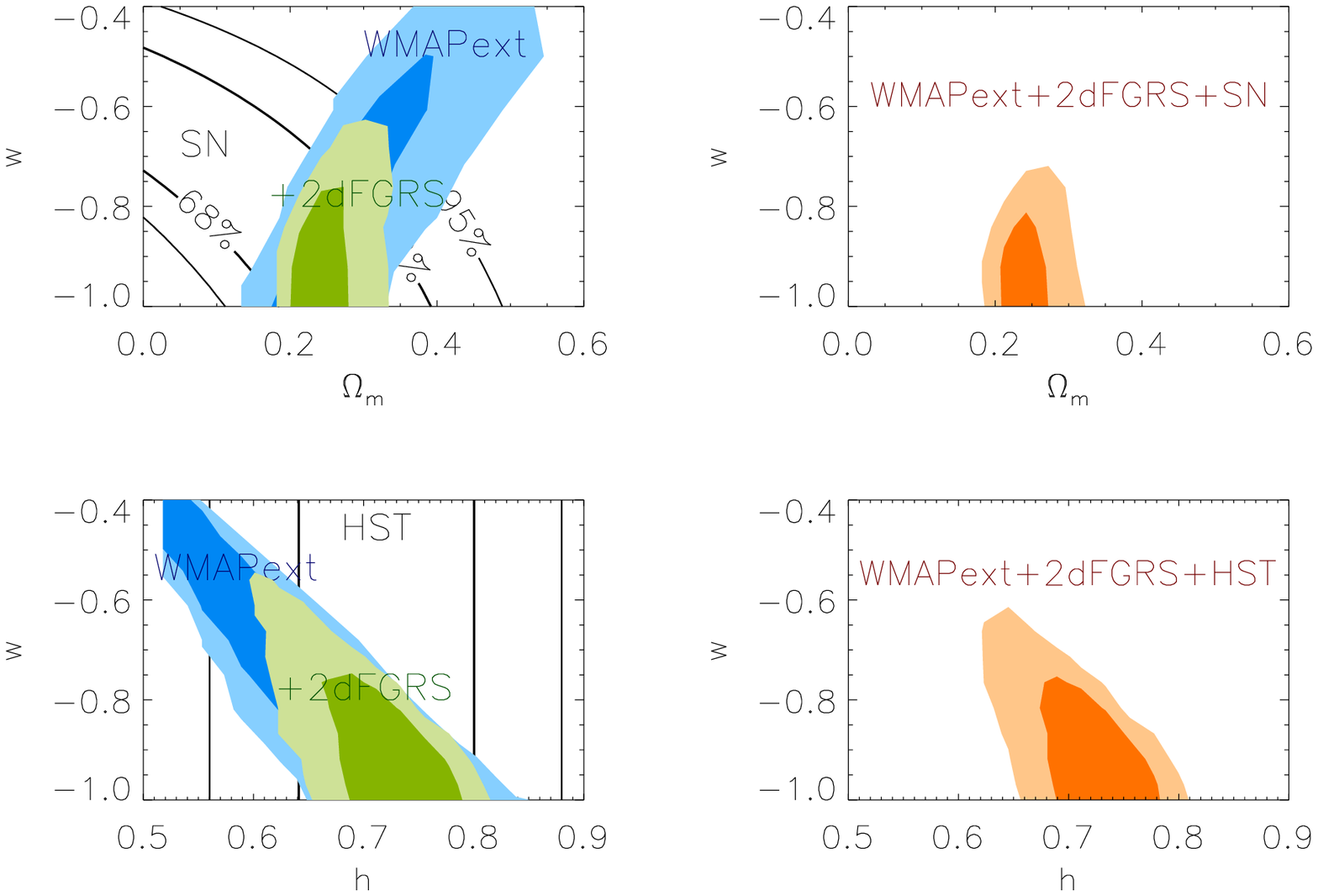}
\caption{Constraints on Dark Energy Properties.   The upper left panel 
shows the marginalized
maximum likelihood surface for the \mapext\  data alone
and for a combination of the \mapext\ + \tdf\ data sets.   The solid
lines in the figure show the 68\% and 95\% confidence ranges 
for supernova data from \citet{riess/etal:2001}.
In the upper right panel, we
multiply  the supernova likelihood function by  the \mapext\ + \tdf\ likelihood
functions. The  lower left
panel shows the maximum likelihood surface for
$h$ and $w$ for the \mapext\ data alone and
 for the \mapext\ + \tdf\ data sets.  The solid lines in
the figures are the 68\% and 95\% confidence limits on $H_0$
from the HST Key Project, where we add the systematic and statistical
errors in quadrature.  In the lower right panel, we multiply the
likelihood function for the \mapext\ + \tdf\ data by the likelihood
surface for the HST data to determine the joint likelihood surface.
The dark areas in these plots are the 68\% likelihood regions and the light
areas are the 95\% likelihood regions. The calculations
for this figure assumed a prior
that $w > -1$.
\label{fig:w_limits}}
\end{figure}

The CMB peak positions constrain the conformal distance to the decoupling
surface.
The amplitude of the early ISW signal determines the
matter density, $\Omega_m h^2$.  The combination of these two measurements
strongly constrains $\Omega(w)$ and $h(w)$ (see Figures~\ref{fig:w}
and \ref{fig:w_limits}).
The HST Key Project measurement of $H_0$ agrees with the inferred
CMB value if $w = -1$.  As $w$ increases, the best fit $H_0$ value for the CMB
drops below the Key Project value.  Our joint analysis of CMB + HST Key Project
data implies that $w < -0.5$ (95\% confidence interval).
If future observations can reduce the uncertainties
associated with the distance to the LMC, the $H_0$ measurements could
place significantly stronger limits on $w$.   Figures~\ref{fig:w} and
\ref{fig:w_limits} show that the combination of
either CMB$+$supernova data or CMB$+$large scale structure data place
similar limits on dark energy properties.
For our combined data set,
we marginalize over all other parameters
and find that \ensuremath{w < -0.78\ (95\%\mbox{\ CL})}  when we
impose the prior that $w > -1$. 
If we drop this prior, then
all of the combined data sets appear to favor a model where
the properties of the dark energy are close to the predicted
properties of a cosmological constant ($w = -0.98 \pm 0.12$).

\subsection{Non-Flat Models}
\label{sec:nf}

The position of the first peak constrains the universe to be nearly flat
\citep{kamionkowski/spergel/sugiyama:1994};
low density models with $\Omega_\Lambda=0$ have their first peak position
at $l \sim 200 \Omega_m^{-1/2}$.  However, if we allow for the possibility that
the universe is non-flat and there is a cosmological constant, then there
is a geometric degeneracy \citep{efstathiou/bond:1999}: along
a line in $\Omega_m -\Omega_\Lambda$ space, there is a set 
of models with  nearly identical angular
power spectra.  While the allowed range of $\Omega_{tot}$ is relatively small,
there is a wide range in $\Omega_m$ values compatible with the CMB data 
in a non-flat universe.

If we place no priors on cosmological parameters, then there is a model
with $\Omega_\Lambda = 0$ consistent with the \map data ($\Delta \chi^2 = 6.6$ 
relative to the flat model).  However,
the cosmological parameters for this model ($H_0 = 32.5$ km/s/Mpc,
and $\Omega_{tot} = 1.28$)
are violently inconsistent with a host of astronomical measurements.
The flat $\Omega_m = 1, \Lambda = 0$ standard CDM model is inconsistent with the 
{\sl WMAP} data at more than the $5 \sigma$ level.

\begin{figure}
\figurenum{13}
\plotone{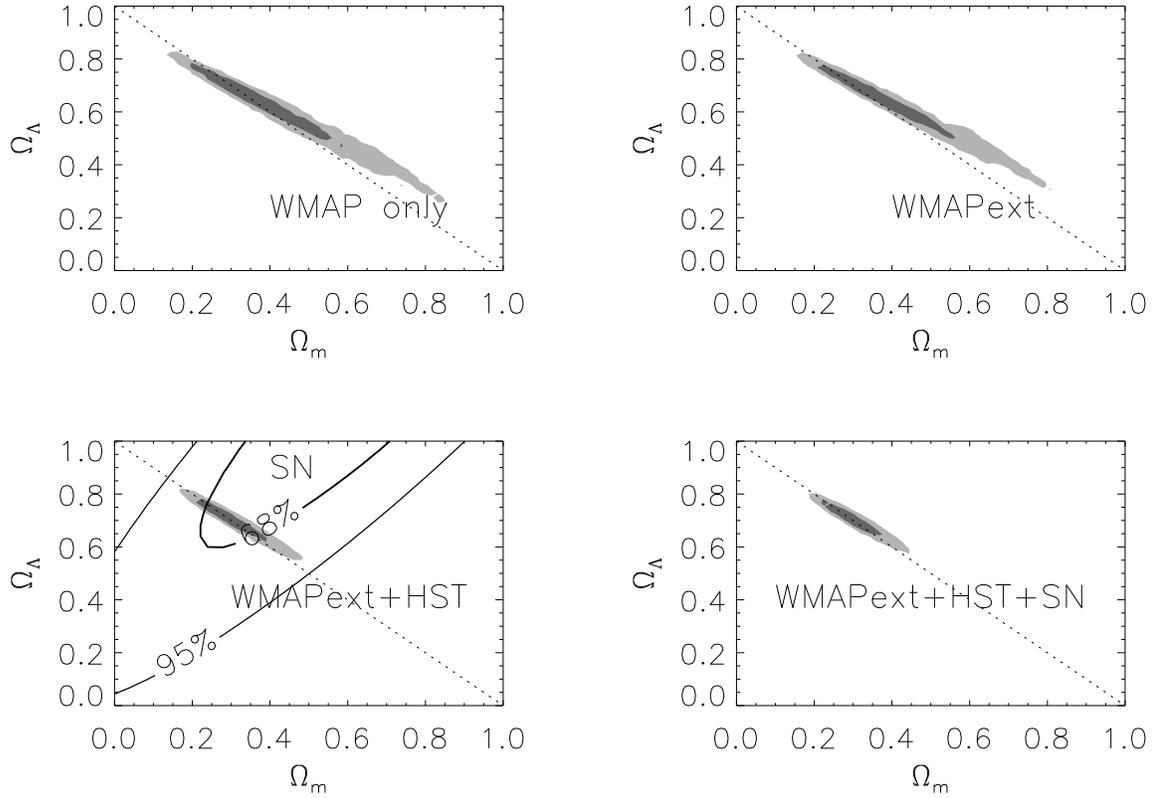}
\caption{Constraints on the geometry of the universe: $\Omega_m-\Omega_\Lambda$ plane.
This figure shows the two dimensional likelihood surface for various combinations
of data: (upper left) \map
(upper right) \mapext
(lower left) \mapext + HST Key Project (supernova data
 \citep{riess/etal:1998,riess/etal:2001} is shown but not used
in the likelihood in this part of the panel; (lower right) \mapext + HST Key Project 
+ supernova 
\label{fig:open}}
\end{figure}

If we include a weak prior on the Hubble Constant, $H_0 > 50 $km/s/Mpc, then
this is sufficient to constrain $0.98 < \Omega_{tot} < 1.08$ (95\% confidence
interval).
Combining the \mapext data with supernova measurements of the angular diameter
distance relationship (see figure \ref{fig:open}) we obtain
$0.98<\Omega_{tot}< 1.06$.   This confidence interval does
not require a prior on $h$.
If we further include the HST Key Project measurement
of $H_0$ as a prior, then the limits on $\Omega_0$ improve slightly:
$\Omega_{tot} < 1.02 \pm 0.02$
Figure \ref{fig:open} shows the two dimensional likelihood surface 
for various combinations of the data.

\subsection{Massive Neutrinos}
\label{sec:mnu}

Copious numbers of neutrinos were produced in the early universe.  If these neutrinos have non-negligible mass they can make a non-trivial
contribution to the total energy density of the universe during both matter and
radiation domination.  During matter domination, the massive neutrinos cluster on
very large scales but free-stream out of smaller scale fluctuations.  This free-streaming
 changes the shape
of the matter power spectrum \citep{hu/eisenstein/tegmark:1998} and most importantly,
suppresses the amplitude of fluctuations.
Since we can normalize the amplitude
of fluctuations to the \map data, the amplitude of fluctuations in the \tdf\ data
places significant limits on neutrino properties.

The contribution of neutrinos to the energy density of the universe depends upon
the sum of the mass of the light neutrino species:
\begin{equation}
\Omega_\nu h^2 = \frac{\sum_i m_i}{94.0 {\rm eV}}.
\end{equation}
Note that the sum only includes neutrino species light enough to decouple while
still relativistic.

Experiments that probe neutrino propagation from source to detector are sensitive
not to the neutrino mass but to the square mass difference between different neutrino
mass eiginstates.
Solar neutrino experiments \citep{bahcall/gonzalez-garcia/penya-garay:2003} imply that the 
square mass difference between
the electron and muon neutrino  is $\sim 10^{-9}$ eV.
The deficit of muon neutrinos in atmospheric showers imply
that the  mass difference between muon and tau neutrinos is
$10^{-5}$eV$^2$ \citep{kearns:2002}.
If the electron neutrino is much lighter
than the tau neutrino, then the combination of these results imply 
that $m_{\nu_\tau} < 0.1$ eV: still below the detection limits for our data-set.
On the other hand,
if $m_{\nu_e} \sim m_{\nu_\tau}$, then the three neutrino
species can leave an observable imprint on the CMB angular
power spectrum and the galaxy large scale structure power spectrum.
In our analysis, we consider this latter case and assume that there
are three degenerate stable light neutrino species.

\begin{figure}
\figurenum{14}
\epsscale{1.0}
\plotone{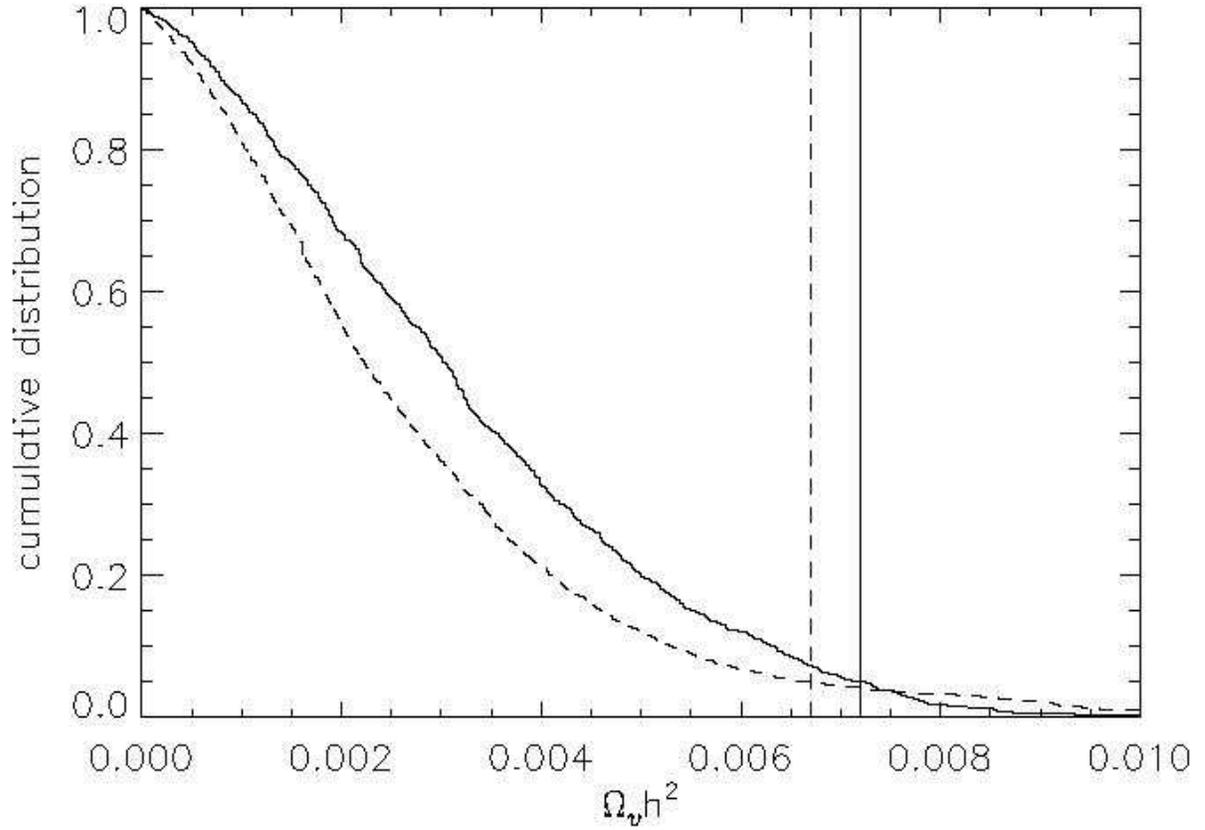}
\caption{This figure shows the marginalized cumulative probability
of $\Omega_\nu h^2$ based on a fit to the \mapext + \tdf\ data sets (dashed)
and the cumulative probability based on a fit to the \mapext + \tdf
+ \lya data sets (solid).  The vertical lines are the 95\% confidence
upper limits for each case (0.21 and 0.23 eV).
\label{fig:mnu}}
\end{figure}

Figure \ref{fig:mnu} shows the cumulative likelihood of the combination of 
{\sl WMAP}, CBI, ACBAR,  and \tdf\ data as a function of the 
energy density in neutrinos.  Based on this analysis, we conclude that 
$\Omega_\nu h^2 < 0.0067$
(95\% confidence limit).
If we add
the \lya data, then the limit slightly weakens to
$\Omega_\nu h^2 < 0.0076$.
For three degenerate neutrino species, this implies
that $m_\nu < 0.23~$eV.  This limit is roughly a factor of two improvement over
previous analyses (e.g., \citet{elgaroy/etal:2002}) that had to
assume strong priors on 
$\Omega_m$ and $H_0$.

\subsection{Tensors}
\label{sec:tensors}

\begin{deluxetable}{llll}
\tablecaption{
\label{tab:tensor}
95\% Confidence Limits on Tensor/Scalar Ratio
}
\tablehead{
\colhead{prior} & \colhead{\map } &
\colhead{\mapext +\tdf }  & \colhead{\mapext + \tdf + \lya }}

\tablewidth{0pt}
\startdata
no prior &
\ensuremath{1.28} &
\ensuremath{1.14} &
\ensuremath{0.90} \\
$dn_s/d\ln k = 0$ &
\ensuremath{0.81} &
\ensuremath{0.53} &
\ensuremath{0.43} \\
$n_s < 1$ &
\ensuremath{0.47} &
\ensuremath{0.37} &
\ensuremath{0.29} \\
\enddata
\end{deluxetable}

Many models of inflation predict a significant gravity wave background.  These tensor
fluctuations were generated during inflation.   Tensor fluctuations have their largest
effects on large angular scales where they add in quadrature to the fluctuations generated
by scalar modes.  

\begin{figure}
\figurenum{15}
\plotone{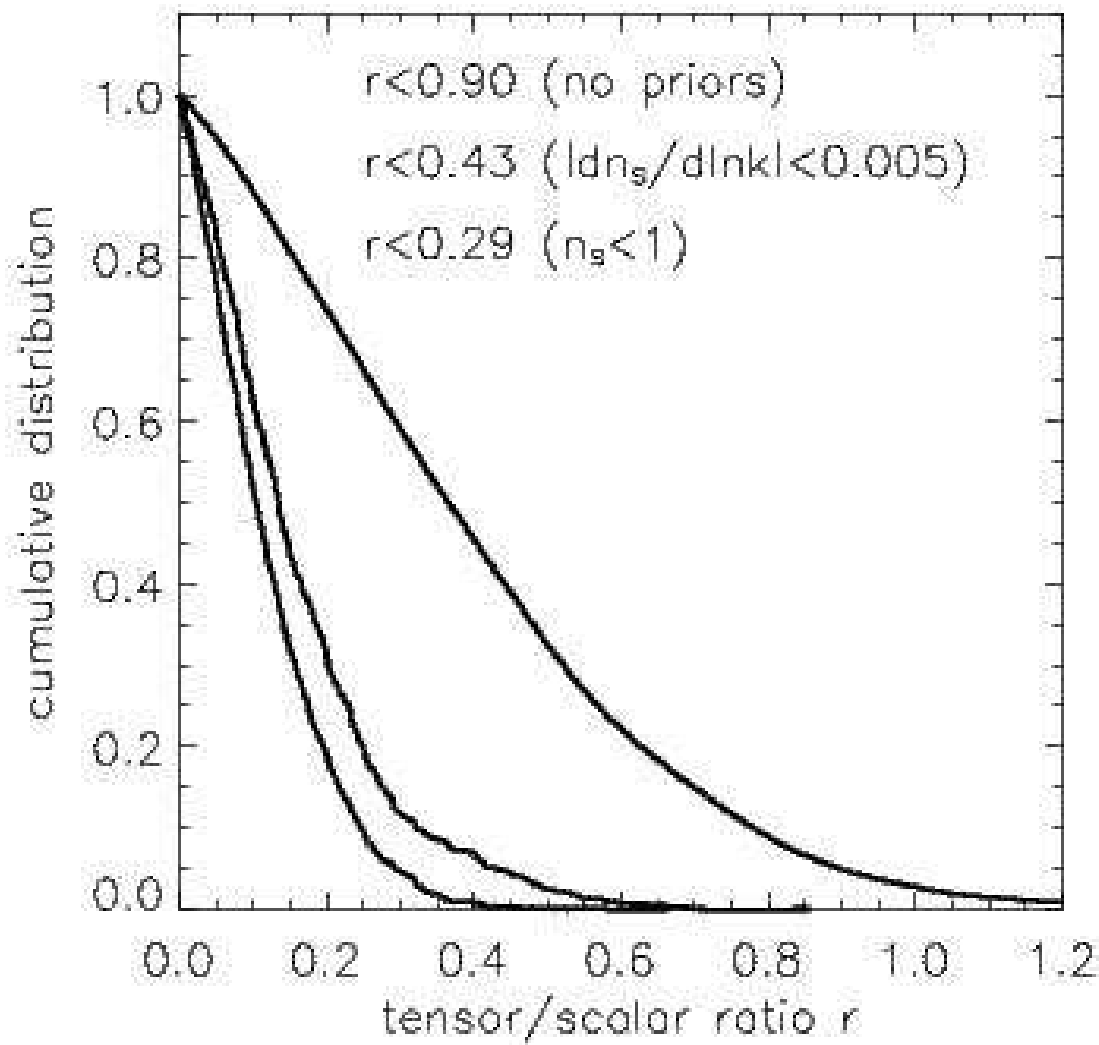}
\caption{This figure shows the cumulative likelihood 
of the combination of the \mapext + \tdf\ +\lya data sets as a function of $r$, the tensor/scalar ratio.  The three lines show the likelihood
for no priors, for models with $\left\vert dn/d\ln k \right\vert
< 0.005$ and for models
with $n_s < 1$.
\label{fig:tensor}}
\end{figure}

Here, we place limits on the amplitude of tensor modes.  We define the tensor
amplitude using the same convention as \citet{leach/etal:2002}:
\begin{equation}
r \equiv \frac{P_{tensor}(k_*)}{P_{scalar}(k_*)},
\end{equation}
where $P_{tensor}$ and $P_{scalar}$ are the primordial amplitude of tensor
and scalar fluctuations and $k_* = 0.002$ Mpc$^{-1}$.  Since we see no evidence
for tensor modes in our fit, we simplify the analysis by assuming that the tensor
spectral index satisfies the single field inflationary
consistency condition:
\begin{equation}
n_{t} = -r/8.
\end{equation}
This constraint reduces the number of parameters in this model to 8:
$A$, $\Omega_b h^2$, $\Omega_m h^2$, $h$, $n_s$,
$d n_s/d\ln k$, $r$ and $\tau$. We ignore the running of $n_t$.
The addition of this new parameter does
not improve the fit as figure (\ref{fig:tensor}) shows the
combination of \mapext + \tdf\  + \lya is able to place a  limit
on the tensor amplitude: $r < 0.90$ (95\% confidence limit).
As table (\ref{tab:tensor}) shows, this limit is much more
stringent if we restrict the parameter space to models with
either $n_s <1$ or $|dn/d\ln k| =0$.

\citet{peiris/etal:2003} discuss the implications of our limits on
tensor amplitude for inflationary scenarios.  Using the results of this
analysis, \citet{peiris/etal:2003} shows that the inferred joint likelihood
of $n_s$, $dn_s/d\ln k$ and $r$ places significant constraints on
inflationary models.

\section{INTRIGUING DISCREPANCIES}
\label{sec:discrepancy}
While the $\Lambda$CDM model's success in fitting CMB data and a host of
other astronomical data is truly remarkable, there remain a pair of intriguing
discrepancies: on both the largest and smallest scales.  While
adding a running spectral index may resolve problems on
small scales, there remains a possible
discrepancy between predictions and observations
on the largest angular scales.

\begin{figure}
\figurenum{16}
\plotone{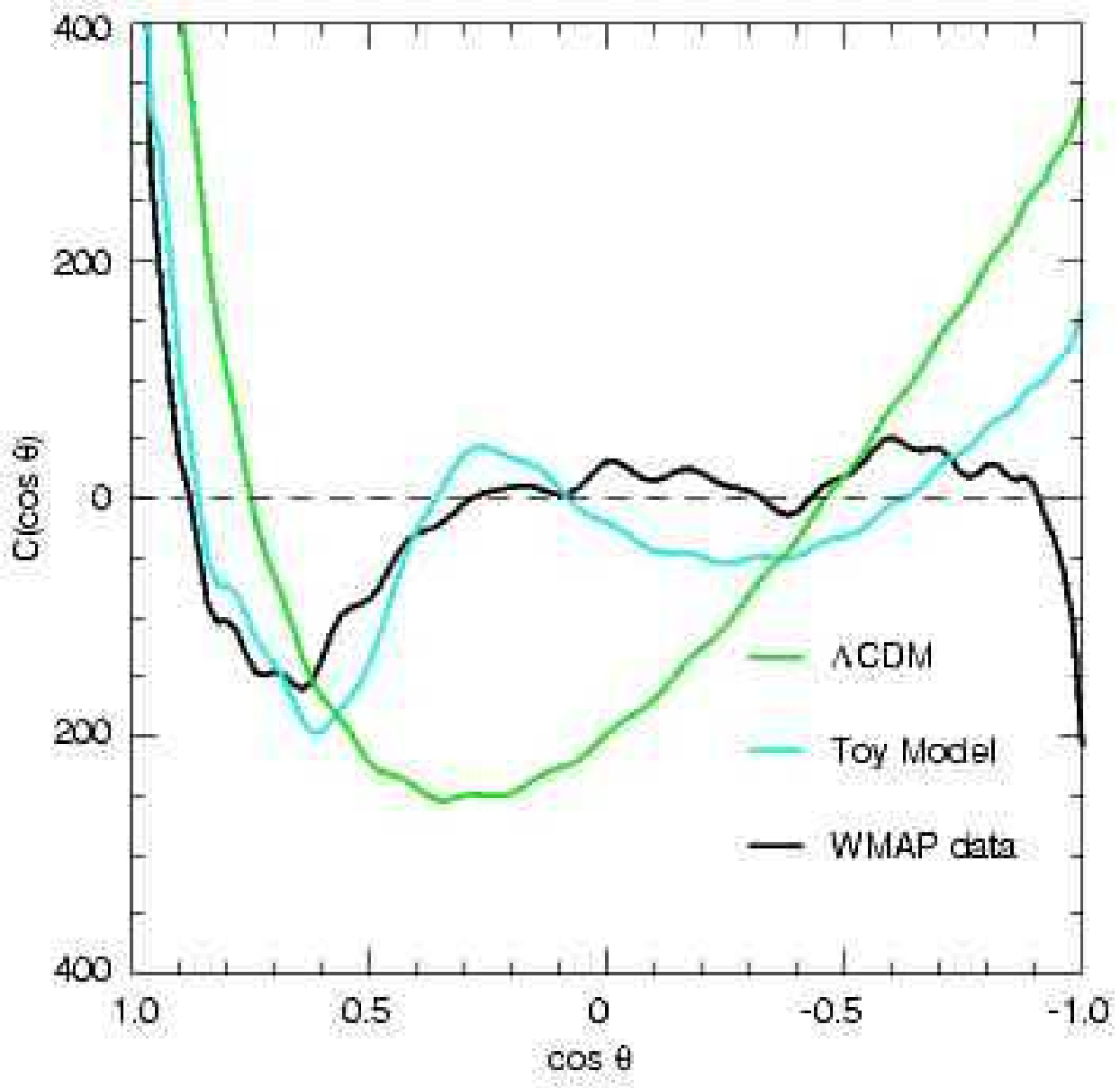}
\caption{Angular correlation 
function of the best fit $\Lambda$CDM model, toy finite
universe model, and \map data on
large angular scales.  The data points are computed from the
template-cleaned V band \map using the Kp0 cut \citep{bennett/etal:2003c}.
\label{fig:toymodel}}
\end{figure}

Figure \ref{fig:toymodel} shows the measured angular power spectrum and 
the predictions of our best fit $\Lambda-$CDM model, where the data were fit
to both CMB and large-scale structure data.  
The figure also shows the measured angular correlation function; the lack of any correlated signal on angular scales  greater than 60 degrees is noteworthy.  
We quantify this lack of power on large scales by measuring a four point 
statistic: 
\begin{equation}
S = \int_{-1}^{1/2} [C(\theta)]^2d\cos\theta.
\end{equation}

The upper cutoff and the form of this statistic were both determined {\it a posteori} in response to the shape of the correlation function.  
We evaluate the statistical significance of these discrepancies by doing Monte-Carlo realizations of the first 100,000 models in the Markov chains.  
This allows us to average not only over cosmic variance but also over our 
uncertainties in cosmological parameters.  For our $\Lambda$CDM Markov 
chains (fit
to the \mapext + \tdf\ data sets), we find that only 0.7\% of the models
have lower values for the quadrupole and only 0.15\% of the simulations
have lower values of $S$.   For the running model, we find that
only 0.9\% of the models have lower  values for the quadrupole
and only 0.3\% of the simulations have lower values of $S$.
The shape of the angular correlation function is certainly 
unusual for realizations of this model.

Is this discrepancy meaningful?  The low quadrupole was already clearly seen in
COBE and was usually dismissed as due to cosmic variance
\citep{bond/jaffe/knox:1998} or foreground contamination.
 While the \map data reinforces the case for its low value, cosmic variance
is significant on these large angular scales and any Gaussian field will always
have unusual features.  On the other hand, 
this discrepancy could be the signature of interesting new physics.  

The discovery of an accelerating universe implies that at these large
 scales, there is new and not understood physics. This new physics is usually interpreted to be dark energy or a 
cosmological constant.  In either case, we would expect that the decay of fluctuations at late times produces a significant ISW signal.   
\citet{boughn/crittenden/turok:1998} argue that in a $\Lambda$CDM model with $\Omega_m = 0.25$, there should be a detectable correlation between the CMB signal and tracers of large-scale structure; yet they were not able to detect a signal. 
There are alternative explanations of the accelerating universe, 
such as extra dimensional gravity theories 
\citep{Deffayet/Dvali/Gabadadze:2002} that do not require a cosmological 
constant and should make radically different predictions for the CMB on these angular scales.  These predictions have not yet been calculated.

\begin{figure}
\figurenum{17}
\plotone{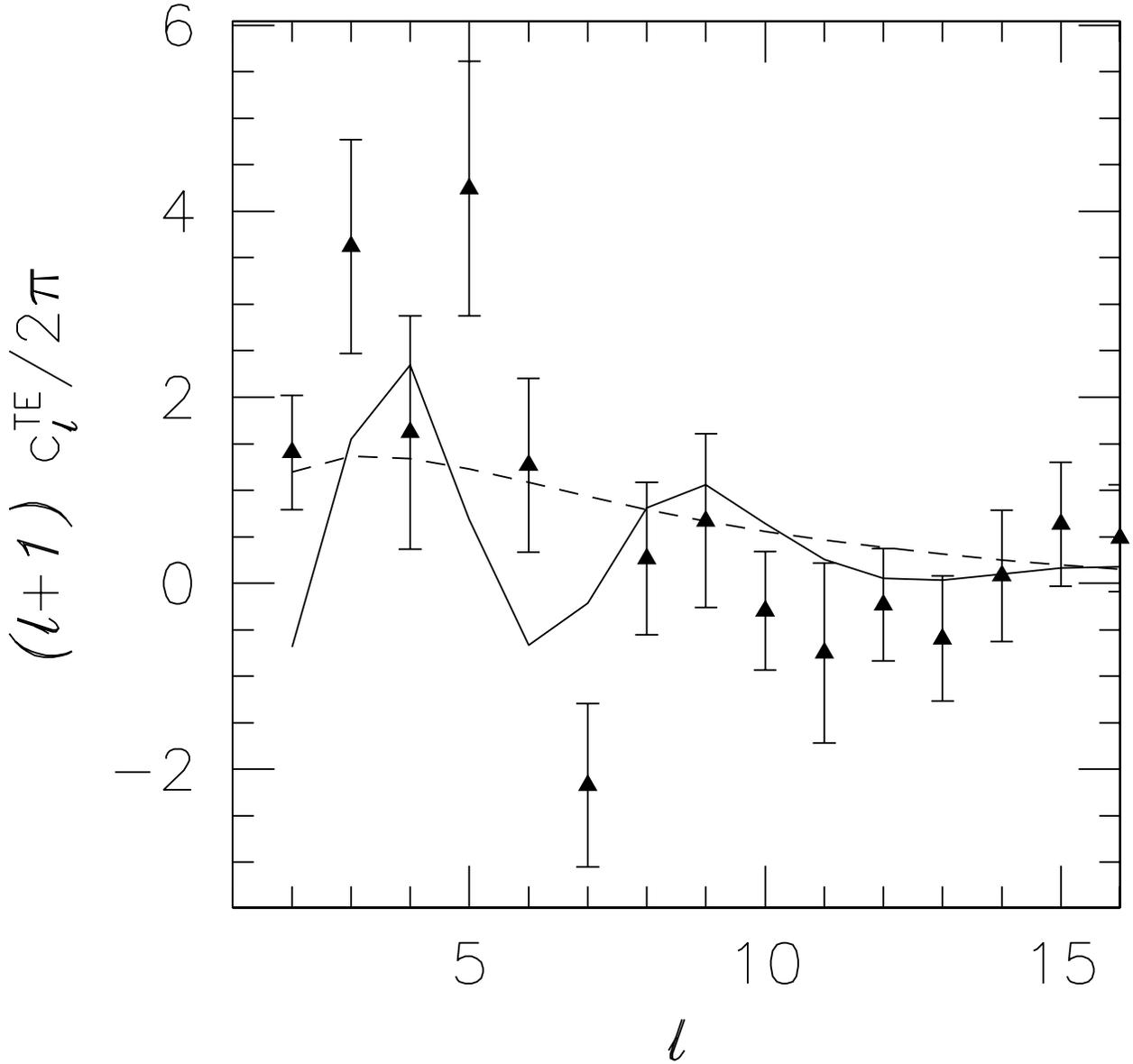}
\caption{TE Power Spectrum.  This figure compares the  data
to
the predicted TE power spectrum
in our toy finite universe model and the $\Lambda$CDM model.
Both models assume that $\tau = 0.17$ and have identical
cosmological parameters.
This figure shows that the TE power spectrum contains
additional information about the fluctuations at
large angles.  While the current data can not distinguish
between these models, future observations could detect the
distinctive TE signature of the model. 
\label{fig:te_toy}}
\end{figure}

What could generate this unusual shaped angular correlation function?
As an example, we compute the angular correlation function in
a toy model, where the power spectrum has the form:
\begin{equation}
P(k) = \sum_{n =1}^{\infty} {\delta(k - 5.8 n/\tau_0) \over k}.
\end{equation}
where $\tau_0$ is the conformal distance to the surface of last scatter.
This toy model simulates both the effects of a discrete power spectrum due to a finite universe and the effects of ringing in the power spectrum due 
to a feature in the inflaton potential (see \citet{peiris/etal:2003}
for a discussion of inflationary models). Figure \ref{fig:toymodel}
shows the angular correlation function and   figure \ref{fig:te_toy} show
the TE power spectrum
of the model.  Note that the TE power spectrum is particularly
sensitive to features in the matter power spectrum.
Intriguingly, this toy model is a better match
to  the observed correlation function than the $\Lambda$CDM model
and predicts a distinctive signature in the TE spectrum.
\citet{cornish/spergel/starkman:1998} show that if the universe
was finite and smaller than the volume within the decoupling surface,
then there should be a very distinctive signal: matched circles.
The surface of last scatter is a sphere centered around {\sl WMAP}.  If the
universe is finite then this sphere must intersect itself,  this leads to pairs
of matched circles.  These match circles provide not only the 
definitive signature
of a finite universe but also should enable cosmologists to 
determine the topology
of the universe\citet{cornish/spergel/starkman:1998b,weeks:1998}.
Should we be able to detect circles if the power spectrum
cutoff is due to the size of the largest mode being $\sim 1/\tau_0$? While
there is no rigorous  theorem relating the size of
the largest mode to the diameter of the fundamental domain, $D$,
analysis of both negatively curved \citep{cornish/spergel:2000}
and positively curved  \citep{lehoucq/etal:2002} topologies suggest
that $D \sim (0.6 - 1) \lambda$.  Thus, if the ``peak'' in the power
spectrum at $l = 5$ corresponds to the largest mode in the domain,
we should be able to detect a pattern of circles in the sky.

Due to the finite size of the patch of the universe visible to \map (or any future satellite), our
ability to determine the origin and significance of this discrepancy will be limited by cosmic variance.  However, future observations can offer some new insight into its origin.
By combining the \map data with tracers of large scale structure
\citep{boughn/crittenden/turok:1998, peiris/spergel:2000}, astronomers may be able to directly detect
the component of the CMB fluctuations due to the ISW effect.  
{\sl WMAP}'s ongoing observations of large-scale microwave background polarization fluctuations will enable additional measurements of fluctuations at large angular scales.  Since the TE observations are probing
different regions of the sky from the TT observations, they
may enlighten  us on whether the lack of correlations on
large angular scales is a statistical fluke or the signature
of new physics.

\section{CONCLUSIONS}

\begin{deluxetable}{ll}
\tablecaption{
Basic and Derived Cosmological Parameters:
Running Spectral
Index Model\tablenotemark{a}
\label{tab:derived_best}}
\tablewidth{0pt}
\tablehead{\colhead{}  &\colhead{Mean and 68\% Confidence Errors}}
\startdata
Amplitude of fluctuations
& \ensuremath{A = 0.83^{+ 0.09}_{- 0.08}} \\
Spectral Index at $k = 0.05$ Mpc$^{-1}$
&\ensuremath{n_s = 0.93 \pm 0.03}\\
Derivative of Spectral Index &
\ensuremath{dn_s/d\ln{k} = -0.031^{+ 0.016}_{- 0.018}} \\
Hubble Constant &
\ensuremath{h = 0.71^{+ 0.04}_{- 0.03}} \\
Baryon Density &
\ensuremath{\Omega_bh^2 = 0.0224 \pm 0.0009}\\
Matter Density & 
\ensuremath{\Omega_mh^2 = 0.135^{+ 0.008}_{- 0.009}}\\
Optical Depth &
\ensuremath{\tau = 0.17 \pm 0.06}\\
\hline
Matter Power Spectrum Normalization &
\ensuremath{\sigma_8 = 0.84 \pm 0.04}\\ 
Characteristic Amplitude of Velocity Fluctuations &
\ensuremath{\sigma_8\Omega_m^{0.6} = 0.38^{+ 0.04}_{- 0.05}}\\
Baryon Density/Critical Density &
\ensuremath{\Omega_b = 0.044 \pm 0.004} \\
Matter Density/Critical Density &
\ensuremath{\Omega_m = 0.27 \pm 0.04}\\
Age of the Universe &
\ensuremath{t_0 = 13.7 \pm 0.2 \mbox{ Gyr}} \\
Reionization Redshift\tablenotemark{b} &
\ensuremath{z_r = 17 \pm 4} \\
Decoupling Redshift &
\ensuremath{z_{dec} = 1089 \pm 1} \\
Age of the Universe at Decoupling & 
\ensuremath{t_{dec} = 379^{+ 8}_{- 7} \mbox{ kyr}}  \\
Thickness of Surface of Last Scatter &
\ensuremath{\Delta z_{dec} = 195 \pm 2} \\
Thickness of Surface of Last Scatter &
\ensuremath{\Delta t_{dec} = 118^{+ 3}_{- 2} \mbox{ kyr}} \\
Redshift of Matter/Radiation Equality &
\ensuremath{z_{eq} = 3233^{+ 194}_{- 210}}\\
Sound Horizon at Decoupling &
\ensuremath{r_s = 147 \pm 2 \mbox{ Mpc}}\\
Angular Size Distance to the Decoupling Surface
&\ensuremath{d_A = 14.0^{+ 0.2}_{- 0.3} \mbox{ Gpc}} \\
Acoustic Angular Scale\tablenotemark{c}
&\ensuremath{\ell_A = 301 \pm 1} \\
Current Density of Baryons &
\ensuremath{n_b = (2.5 \pm 0.1) \times 10^{-7} \mbox{ cm$^{-3}$}} \\
Baryon/Photon Ratio &
\ensuremath{\eta = (6.1^{+ 0.3}_{- 0.2}) \times 10^{-10} \mbox{ }} \\
\enddata
\tablenotetext{a}{Fit  to the \map, CBI, ACBAR, \tdf\ and \lya forest data}
\tablenotetext{b}{Assumes ionization fraction, $x_e = 1$}
\tablenotetext{c}{ $l_A=\pi d_C/r_s$}
\end{deluxetable}

Cosmology now has a standard model:
a flat universe composed of matter, baryons and vacuum energy with a nearly scale-invariant spectrum
of primordial fluctuations.  In this 
cosmological model, the properties
of the universe are characterized
by the density of baryons, matter and the expansion rate: $\Omega_b$, 
$\Omega_m,$ 
and $h$.  For the analysis of CMB results, all of the effects of
star formation can be incorporated in a single number:
the optical depth due to reionization, $\tau$.  
The primordial fluctuations in this model
are characterized by a spectral index.  
Despite its simplicity, it
 is an adequate fit not only to the
\map temperature and polarization data but also to small scale CMB data, large scale structure
data, and supernova data.  This model is consistent with the baryon/photon ratio
inferred from observations of $D/H$ in distant quasars, the HST Key
Project measurement of the Hubble constant, stellar  ages and
the amplitude of mass fluctuations inferred from clusters and from gravitational lensing.  When we include large scale structure or Lyman $\alpha$ forest data in the analysis,
the data suggest that we may need to add an additional parameter:
$dn_s/d\ln k$. Since the best fit models predict that the 
slope of the power spectrum is redder on small scales, this model
predicts later formation times for dwarf galaxies.  This modification
to the power law $\Lambda$CDM model may resolve many of its problems on
the galaxy scale.
Table (\ref{tab:derived_best}) lists the
best fit parameters for this model.

While there have been a host of papers on cosmological parameters, \map has
brought this program to a new stage: \map's more accurate determination of
the angular power spectrum has significantly reduced parameter uncertainties,
\map's detection of TE fluctuations has confirmed the
basic model and its detection of reionization signature has reduced the $n_s-\tau$
degeneracy. Most importantly, the rigorous propagation of errors
and uncertainties in the \map data has strengthened the significance of
the inferred parameter values.

In this paper, we have also examined a number of more complicated 
models: non-flat universes, quintessence models,  
models with massive neutrinos, and models with tensor gravitational wave
modes.
By combining the \map data with finer scale CMB experiments and
with other astronomical data sets (\tdf\ galaxy power spectrum and SNIa observations),
we place significant new limits on these parameters. 

Cosmology is now in a similar stage in its intellectual development to
particle physics three decades ago when particle physicists converged
on the current standard model.  The standard model of particle physics fits
a wide range of data, but does not answer many fundamental
questions: ``what is the origin of mass? why is there more than
one family?, etc."  
Similarly, the standard cosmological model has many deep open 
questions: "what is the dark energy? what is the dark matter? what
is the physical model behind
inflation (or something like inflation)?" 
Over the past three decades, precision
tests have confirmed the standard model of
particle physics and searched for distinctive
signatures of the natural extension of the standard model: supersymmetry.
Over the coming years, improving CMB, large scale structure, lensing,
and supernova data will provide ever more rigorous tests of the cosmological
standard model and search for new physics beyond the standard model.

\section*{ACKNOWLEDGMENTS}

We thank Ed Jenkins for helpful comments about the [D/H]  recent measurements and their interpretation. We thank Raul Jimenez for useful discussions 
about the cosmic ages. We thank Adam Riess for providing us with the likelihood surface form the SNIA data. The \map mission is made possible by the support of the Office Space at NASA Headquarters and by the hard and capable work of scores of scientists, engineers, managers, administrative staff, and reviewers.
LV is supported by NASA through Chandra Fellowship PF2-30022 issued
by the Chandra X-ray Observatory center, which is operated by
the Smithsonian Astrophysical Observatory for and on behalf of
NASA under contract NAS8-39073.

\clearpage

\begin{figure}
\figurenum{8}
\plotone{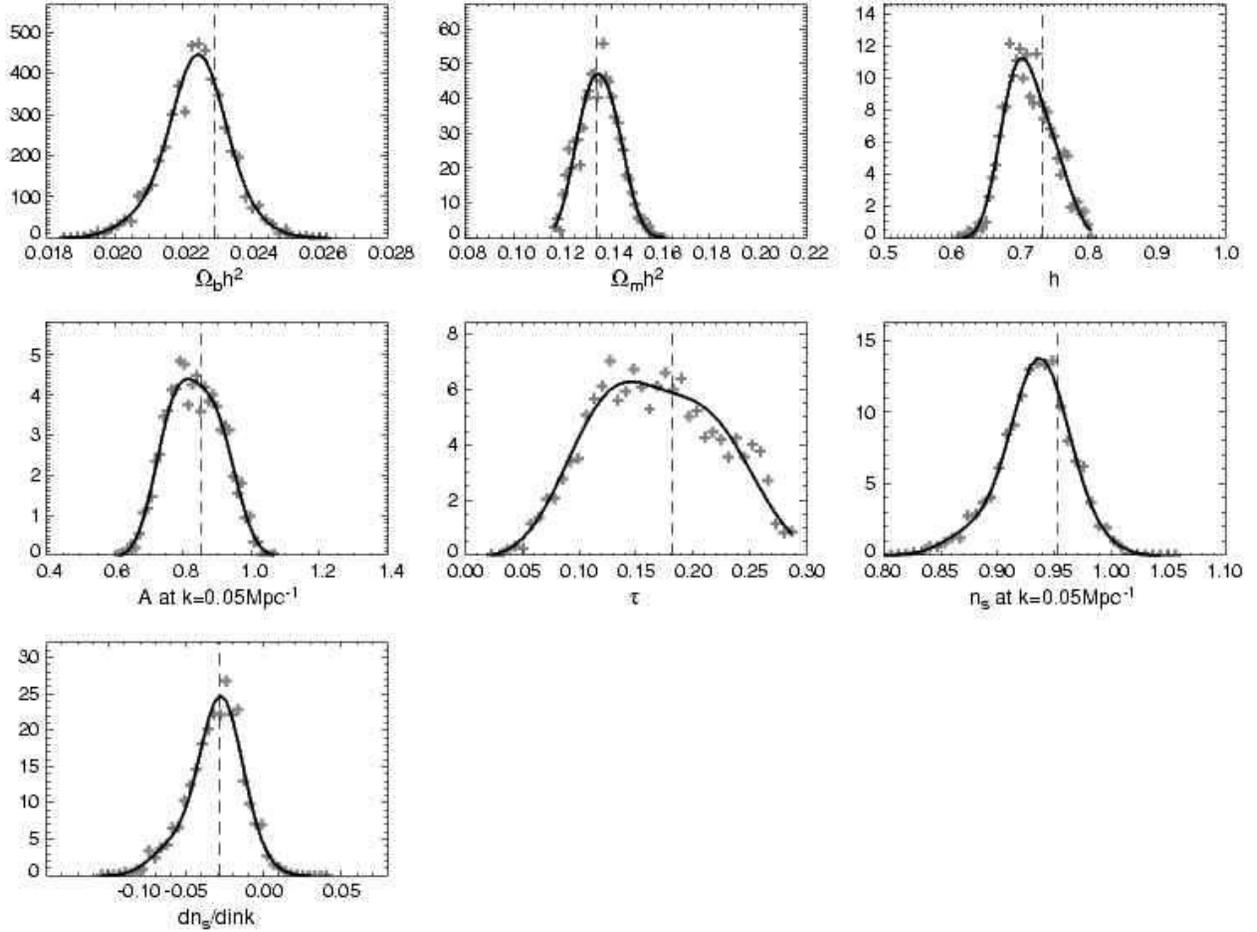}
\caption{This figure shows the marginalized likelihood for
various cosmological parameters in the running
spectral index model for our analysis of the combined
\map, CBI, ACBAR, 2dFGRS and Lyman $\alpha$ data sets.
The dashed lines show the maximum likelihood values of the global
seven dimensional fit.
\label{fig:dndlnk}}
\end{figure}

\end{document}